\documentclass[useAMS,usenatbib]{mn2e}
\usepackage{graphicx}
\usepackage{fixltx2e}
\usepackage{enumerate}
\usepackage{rotating}
\usepackage{multirow}
\usepackage{url}
\title[A high resolution study of COMs in hot cores] {A high resolution study of complex organic molecules in hot cores}
\author[Calcutt et al.]{Hannah Calcutt$^{1}$\thanks{E-mail: hcalcutt@star.ucl.ac.uk}, 
Serena Viti$^{1}$, Claudio Codella$^{2}$, Maria T. Beltr\'{a}n$^{2}$, Francesco Fontani$^{2}$, \newauthor Paul M. Woods$^{1,3}$\\
$^{1}$ Department of Physics and Astronomy, University College London, WC1E 6BT, London, UK\\
$^{2}$ INAF, Osservatorio Astrofisico di Arcetri, Largo Enrico Fermi 5, I-50125 Firenze, Italy\\ 
$^{3}$ Astrophysics Research Centre, School of Mathematics \& Physics, Queen's University Belfast, Belfast BT7 1NN, UK\\}

\begin{document}

\date{}

\pagerange{\pageref{firstpage}--\pageref{lastpage}} \pubyear{2013}

\maketitle

\label{firstpage}

\begin{abstract}
We present the results of a line identification analysis using data from
the IRAM Plateau de Bure Inferferometer, focusing on six massive star-forming
hot cores: G31.41+0.31, G29.96-0.02, G19.61-0.23, G10.62-0.38, G24.78+0.08A1 and G24.78+0.08A2.
 We identify several transitions of
vibrationally excited methyl formate (HCOOCH$_3$) for the first time in these
objects as well as transitions of other complex molecules, including
ethyl cyanide
(C$_2$H$_5$CN), and isocyanic acid (HNCO).  We also postulate a detection of one transition of glycolaldehyde (CH$_2$(OH)CHO) in two new hot cores. We find G29.96-0.02, G19.61-0.23, G24.78+0.08A1 and G24.78+0.08A2 to be chemically very similar. G31.41+0.31,
however, is chemically different: it manifests a larger chemical inventory
and has significantly larger column densities. We suggest that it may
represent a different evolutionary stage to the other hot cores in
the sample, or it may surround a star with a higher mass. We derive column
densities for methyl formate in G31.41+0.31, using the rotation diagram
method, of 4$\times$10$^{17}$ cm$^{-2}$ and a T$_{rot}$ of $\sim$170 K.   For G29.96-0.02, G24.78+0.08A1 and G24.78+0.08A2,
 glycolaldehyde, methyl formate and methyl cyanide all seem to trace the same material and peak at roughly the same position towards the dust emission peak. For G31.41+0.31, however, glycolaldehyde shows a different distribution to methyl formate and methyl cyanide and seems to trace the densest, most compact inner part of hot cores.
\end{abstract}


\def\HII{H{\sc ii }}
\def\aj{Astronomical Journal}
\def\actaa{Acta Astronomica}
\def\araa{Annual Review of Astron and Astrophys}
\def\apj{Astrophysical Journal}
\def\apjl{Astrophysical Journal, Letters}
\def\apjs{Astrophysical Journal, Supplement}
\def\ao{Applied Optics}
\def\apss{Astrophysics and Space Science}
\def\aap{Astronomy and Astrophysics}
\def\aapr{Astronomy and Astrophysics Reviews}
\def\aaps{Astronomy and Astrophysics, Supplement}
\def\azh{Astronomicheskii Zhurnal}
\def\baas{Bulletin of the AAS}
\def\caa{Chinese Astronomy and Astrophysics}
\def\cjaa{Chinese Journal of Astronomy and Astrophysics}
\def\icarus{Icarus}
\def\jcap{Journal of Cosmology and Astroparticle Physics}
\def\jrasc{Journal of the RAS of Canada}
\def\memras{Memoirs of the RAS}
\def\mnras{Monthly Notices of the RAS}
\def\na {New Astronomy}
\def\nar{New Astronomy Review}
\def\pra{Physical Review A: General Physics}
\def\prb{Physical Review B: Solid State}
\def\prc{Physical Review C}
\def\prd{Physical Review D}
\def\pre{Physical Review E}
\def\prl{Physical Review Letters}
\def\pasa{Publications of the Astron. Soc. of Australia}
\def\pasp{Publications of the ASP}
\def\pasj{Publications of the ASJ}
\def\rmxaa{Revista Mexicana de Astronomia y Astrofisica}
\def\qjras{Quarterly Journal of the RAS}
\def\skytel{Sky and Telescope}
\def\solphys{Solar Physics}
\def\sovast{Soviet Astronomy}
\def\ssr{Space Science Reviews}
\def\zap{Zeitschrift fuer Astrophysik}
\def\nat{Nature}
\def\iaucirc{IAU Cirulars}
\def\aplett{Astrophysics Letters}
\def\apspr{Astrophysics Space Physics Research}
\def\bain{Bulletin Astronomical Institute of the Netherlands}
   \def\fcp{Fundamental Cosmic Physics}
   \def\gca{Geochimica Cosmochimica Acta}
   \def\grl{Geophysics Research Letters}
   \def\jcp{Journal of Chemical Physics}
   \def\jgr{Journal of Geophysics Research}
   \def\jqsrt{Journal of Quantitiative Spectroscopy and Radiative Transfer}
   \def\memsai{Mem. Societa Astronomica Italiana}
   \def\nphysa{Nuclear Physics A}
   \def\physrep{Physics Reports}
   \def\physscr{Physica Scripta} 
   \def\planss{Planetary Space Science}
   \def\procspie     {Proceedings of the SPIE}
   
\section{Introduction}

The process that leads to the formation of a high-mass star, whether
it is by accretion \citep{mckee2002,mckee2003}, or coalescence and
accretion \citep{bonnell1998} is extremely fast,
$\sim$10$^4$--$10^5$\,yr, and will, according to some evolutionary models,
depend on the final mass of the star \citep[e.g.,][]{bernasconi1996}.
The contraction of a core from a low to a very high density
($\sim$10$^7$\,cm$^{-3}$) occurs very quickly and the star will reach
the Zero Age Main Sequence while still embedded \citep{palla1993}.
This, together with the fact that massive stars tend to form in
association, makes the determination of the early stages of the
evolution of a massive star via observations of Spectral Energy
Distributions (SEDs) a rather difficult task.

A common way of observing massive young stellar objects (YSOs) has
been via the detection of hypercompact and ultracompact HII regions
(UCHII): regions of dense gas ionized by the newly formed star.
However, ionised regions trace relatively late evolutionary stages. On
the other hand, the earliest phases are well traced by hot cores,
which are small ($\sim$10$^{-2}$--$10^{-1}$\,pc), dense
($\geq$10$^7$\,cm$^{-3}$), relatively warm ($\geq$10$^2$\,K),
optically thick (A$_v$$\geq$10$^2$\,mag), and transient
($\leq$10$^5$\,yr) condensations. Spectral surveys have revealed a
very rich chemistry (see review by \citet{herbst2009,garay1999}) which include high
abundances of small saturated molecules (e.g. H$_2$O, NH$_3$, H$_2$S,
CH$_3$OH) as well as large organic species (CH$_3$CN, CH$_2$CHCN,
CH$_3$CH$_2$CN, C$_2$H$_5$OH).  Such high abundances are believed to
arise from the sublimation of ice mantles frozen out onto the dust
grains during the collapse of the parent cloud, which is induced by
the nearby (proto)star.

Once the protostar is born, the dust is heated and molecules formed on
the dust grain sublimate. This warm-up phase is most likely not
instantaneous \citep{viti1999}; \citep{viti2004}; \citep{garrod2008}, and there is experimental evidence that desorption
occurs in as many as four stages in four distinct narrow temperature ranges (temperature bands). The proportion of each
species that evaporates in each band depends on the total amount
present on the grains as well as on the bonding properties of the molecule
\citep{collings2004}.  Chemical models incorporating such experimental
results show that distinct chemical events occur at specific grain
temperatures and these differ depending on the mass of the star
\citep{viti2004}. The chemistry of the hot cores will, therefore,
reflect this temperature-driven evolution and the evolutionary
chemistry can provide, in principle, a record of the collapse process
as well as the ignition history of the star.

Of course, since there is a chemical differentiation over time due to
a time-dependent increase of the dust temperature, one also expects a
space-dependent chemical differentiation due to the difference in dust
temperature progressively further away from the star. The degeneracy
between time and spatial effects can only be solved by interferometric
observations and chemical modelling.  A class of molecules that can be
particularly useful to solve this degeneracy is that of complex
organic molecules (COMs) because (i) their detection is a confirmation
of the high density warm cores, as most COMs are not easily produced
in the gas phase chemistry of dark clouds and, ii) most importantly,
their emission has been observed to be compact in extent in star forming regions outside the Galactic Center
\citep{herbst2009,garay1999} and therefore may trace the most central
region of the hot core, close to the YSOs.  In fact, many organic
molecules have been detected in hot cores and have been used to trace
different structures associated with star formation such as discs and
maser activity
\citep[e.g.][]{olmi1996,cesaroni1998,olmi2003,beltran2005,beltran2011}.
Multi-transition and multi-species observations of complex molecules
are essential in order to derive the temperature and excitation
conditions of hot cores, especially if different transitions and
species trace different extensions of the core.

\citet{beltran2005} mapped the CH$_3$CN emission in two hot cores and
spectrally identified many other organic molecules in both
objects. This work led to the first detection outside the Galactic
Centre of a transition of glycolaldehyde, CH$_2$(OH)CHO, in
G31.41+0.31 \citep{beltran2009}. The richness of these spectra gives a
great insight into the chemistry of this type of object, which we
investigate in depth here. In addition to the previously-published
data for G31.41+0.31 and G24.78+0.08 A2, we also analyse the
unpublished spectra of four more hot cores to give a sample of six
(see Sect.~\ref{sect:obs}). Seven emission lines evaded definite
identification in the spectrum of G31.41+0.31 at 1.4\,mm. In this
paper we identify some of these unidentified lines, and find that
they are seen in other members of our sample. Since all the new
identifications are of complex species, we use the new data in
conjunction with previously-published data to derive the excitation
conditions of each core. We then make use of a chemical model of a hot
core to interpret the molecular inventory of the six cores and
qualitatively characterise each core and its evolutionary stage.  Previous work such as \citet{isokoski2013} have taken a molecular inventory of hot cores using single dish observations to look for chemical differences in objects with disk-like structures compared to those without disks. This work will explore the chemical differences between hot cores by making use of high spatial resolution observations. This will allow us to look at the spatial chemical variation in hot cores as well as the variability between objects. 

\section{Observations} 
\label{sect:obs}
This work is based on observations taken with the Plateau de Bure Interferometer (PdBI) and reported by \citet{beltran2005}, for the hot cores G31.41+0.31 (G31) and A1 and A2 in G24.78+0.08 (G24A1 and G24A2), and by \citet{beltran2011} for the hot cores G29.96-0.02 (G29), G19.61-0.23 (G19), G10.62-0.38 (G10). Information about the observations can be found in their Section 2. The synthesized CLEANed beams for maps made using natural weighting can be found in Table \ref{tab:beamsize}. The V$_{LSR}$ for each hot core listed in Table \ref{tab:beamsize} has been determined from high spatial resolution observations of the 12-13 transitions of CH$_3$$^{13}$CN.

\begin{table*}
\caption{Parameters of the IRAM PdBI observations\label{tab:beamsize}}
\begin{center}
\scriptsize{
\begin{tabular}{ccccccc}
\hline

Source & $\alpha$(J2000)$^{\dagger}$&$\delta$(J2000)&V$_{LSR}$  & Synthesized Beam  & P.A.&Velocity resolution (kms$^{-1}$) \\
       &(h m s)&($\degr \arcmin \arcsec$)&(kms$^{-1}$)&$^{\dagger\dagger}$ ($\arcsec$) &($\degr$)\\
\hline
G31.41+0.31&18 47 34.330&-01 12 46.50&\phantom{1}96.8$^a$&1.1 $\times$ 0.5&-170&3.4\\
G29.96-0.02&18 46 03.955&-02 39 21.87&\phantom{1}98.9$^b$&1.4 $\times$ 0.7&\phantom{-}168&3.4\\
G19.61-0.23&18 27 38.145&-11 56 38.49&\phantom{1}41.6$^b$& 2.6 $\times$ 1.0&\phantom{-}161&3.4\\
G10.62-0.38&18 10 28.650&-19 55 49.50&\phantom{1}-2.0$^b$&2.4 $\times$ 0.7&\phantom{-}169&3.4\\
G24.78+0.08&18 36 12.660&-07 12 10.15&110.8$^a$&1.2 $\times$ 0.5&-174&3.4\\

\hline

\end{tabular}

}
\end{center}
\begin{flushleft}
{\raggedright $^{\dagger}$ Coordinates of the phase centre of the observations. \\$^{\dagger\dagger}$ The synthesized CLEANed beams for maps made using natural weighting. \\
 $^a$\citet{beltran2005} \\
 $^b$\citet{beltran2011}}
\end{flushleft}
\end{table*}

These observations range from 220\,209.95\,MHz to 220\,759.69\,MHz for all the hot cores and were analysed using the
GILDAS software package. \footnote{http://www.iram.fr/IRAMFR/GILDAS}

\section{The sample of hot cores}

We have selected luminous ($L_\mathrm{bol}>10^4$\,L$_\odot$) objects
with typical signposts of massive star formation such as water masers
and ultracompact (UC) HII regions. 

Six hot cores were observed towards five star-forming regions on the
basis of being bright in the sub-mm,  and having previously given indications of being chemically
rich.  We discuss each source in detail below.

{\it G31.41+0.31:} This is a well-studied hot core located at 7.9\,kpc
\citep{cesaroni1994,cesaroni1998}, with evidence of a rotating massive
molecular toroid, suggested by OH maser emission and confirmed using
CH$_3$CN emission \citep{gaume1987,beltran2004,beltran2005,cesaroni2011}.  It has been
mapped in several different molecules including SiO, HCO$^+$ and
NH$_3$, as well as several complex molecules like CH$_3$CN,
C$_2$H$_5$CN and CH$_2$(OH)CHO
\citep{cesaroni1994,maxia2001,beltran2005}.  Interferometric
observations of molecular lines with high excitation energies have
revealed the presence of deeply-embedded YSOs, which in all likelihood
explains the temperature increase toward the core centre
\citep{beltran2004,beltran2005,cesaroni2010}. The Spitzer/GLIMPSE
images by \citet{benjamin2003} show that the G31 hot core lies in a
complex parsec-scale region where both extended emission and multiple
stellar sources are detected.

{\it G29.96-0.02:} G29 is located at a distance of 3.5\,kpc
\citep{beltran2011} and associated with the infrared source IRAS
18434-0242. It contains a cometary UCHII region with a hot core
located in front of the cometary arc \citep{wood1989,cesaroni1994}.
It has been mapped in several molecules including NH$_{3}$, HCO$^+$,
CS, CH$_{3}$CN, HNCO and HCOOCH$_3$
\citep{cesaroni1998,pratap1999,maxia2001,olmi2003,beuther2007}. A
velocity gradient along the east-west direction has been measured in
NH$_3$, CH$_3$CN, and HN$^{13}$C emission which is interpreted as
rotation \citep{cesaroni1998,olmi2003,beuther2007}.  On the other
hand, an outflow directed along the southeast-northwest direction has
been mapped in H$_2$S and SiO by \citet{gibb2004} and
\citet{beuther2007}. \citet{beltran2001} confirmed the existence of a rotating
molecular toroid around the outflow axis.  Masers of H$_2$O, CH$_3$OH
and H$_2$CO have also been detected around this hot core
\citep{hofner1996,walsh1998,hoffmann2003}.

{\it G19.61-0.23:} G19 is located at a distance of 12.6\,kpc
\citep{kolpak2003}, and associated with the infrared source IRAS
18248-1158. It contains several embedded UCHII regions, detected by
\citet{garay1985}, and more recently mapped by
\citet{furuya2005}. Several molecular tracers, such as CS, NH$_3$,
CH$_3$CH$_2$CN, HCOOCH$_3$, and CH$_3$CN have been mapped in this hot
core \citep{plume1992,garay1998,remijan2004,furuya2005}.  $^{13}$CO
and C$^{18}$O emission show inverse P Cygni profiles indicating
infalling gas towards the core \citep{wu2009,furuya2011}.
\citet{lopez2009} mapped a $^{13}$CO outflow without a well-defined
morphology. H$_2$O, OH, and CH$_3$OH masers have been detected by
\citet{forster1989}, \citet{hofner1996} and \citet{walsh1998}.
\citet{beltran2011} reported velocity gradients, observed in CH$_3$CN,
oriented perpendicular to the direction of a molecular outflow.

{\it G10.62-0.38:} G10 is located at a distance of 3.4\,kpc
\citep{blum2001} and contains a well-studied UCHII region
\citep[e.g.][]{wood1989} associated with the infrared source IRAS
18075-1956.  The hot core in this star forming region has been mapped
in NH$_3$ \citep{ho1986,keto1987,keto1988,sollins2005} and in SO$_2$
and OCS \citep{klaassen2009}.  Infall and bulk rotation in the
molecular gas surrounding the UCHII region have been detected.
H66$\alpha$ emission shows the occurrence of inward motions in the
ionised gas \citep{keto2002}.  CH$_3$OH and H$_2$O masers have been
mapped towards the core and are distributed linearly in the plane of
the rotation \citep{hofner1996,walsh1998}, while OH masers appear to
lie along the axis of rotation \citep{argon2000}. Outflow activity has
been detected by \citet{keto2006}, \citet{lopez2009} and by
\citet{beltran2011}, who reported a CH$_3$CN toroid rotating around
the main axis of the outflow.

{\it G24.78+0.08A1 and G24.78+0.08A2:} G24 is a high-mass star-forming
region located at 7.7\,kpc from the Sun, and associated with several
YSOs in different evolutionary phases embedded in their parental cores
\citep{furuya2002}.  The G24 region has been extensively studied in
various molecules like CO, NH$_3$ and CS, as well as in complex
molecules and in the continuum
\citep{codella1997,cesaroni2003,beltran2005,beltran2011}.  Two main
groups of YSOs, called A1 and A2, are separated by $\sim$1\farcs5
\citep[][and references therein]{beltran2011}.  {\it G24.78+0.08A1:}
G24A1 is one of the three massive cores with a rotating toroid detected in G24 \citep{beltran2011}.  At the centre of
G24A1, an unresolved hypercompact (HC) HII region is created by a YSO
with a spectral type of at least O9.5 \citep{codella1997,beltran2007}.
NH$_3$ (2,2) observations have revealed that the gas in the toroid is
undergoing infall towards the HC HII region \citep{beltran2006},
suggesting that accretion onto the star might still be occurring, even
through the ionised region \citep[for a potential mechanism,
  see][]{keto2006}.  On the other hand, Very Long Baseline Array
proper motion measurements of H$_2$O masers associated with the HC HII
region \citep{moscadelli2007} indicate that the ionised region might
be expanding, thus questioning the possibility of accretion onto the
star.  {\it G24.78+0.08A2:} G24A2 is also associated with a massive
CH$_3$CN toroid, rotating around the main axis of a bipolar outflow
observed in the CO isotopologues
\citep{furuya2002,beltran2005,beltran2011,codella2013}. The mid-infrared and radio
continuum measurements show a compact (1000--2000\,AU) source
which could be due to an ionised jet \citep{vig2008}.  Emission from
several complex molecules clearly indicate the presence of a molecular
hot core \citep{beltran2005,beltran2011, codella2013}.

\section{Molecular content}
\label{sec:molcon}

\begin{figure*}
\begin{center}
\includegraphics[angle=-90,width=15cm]{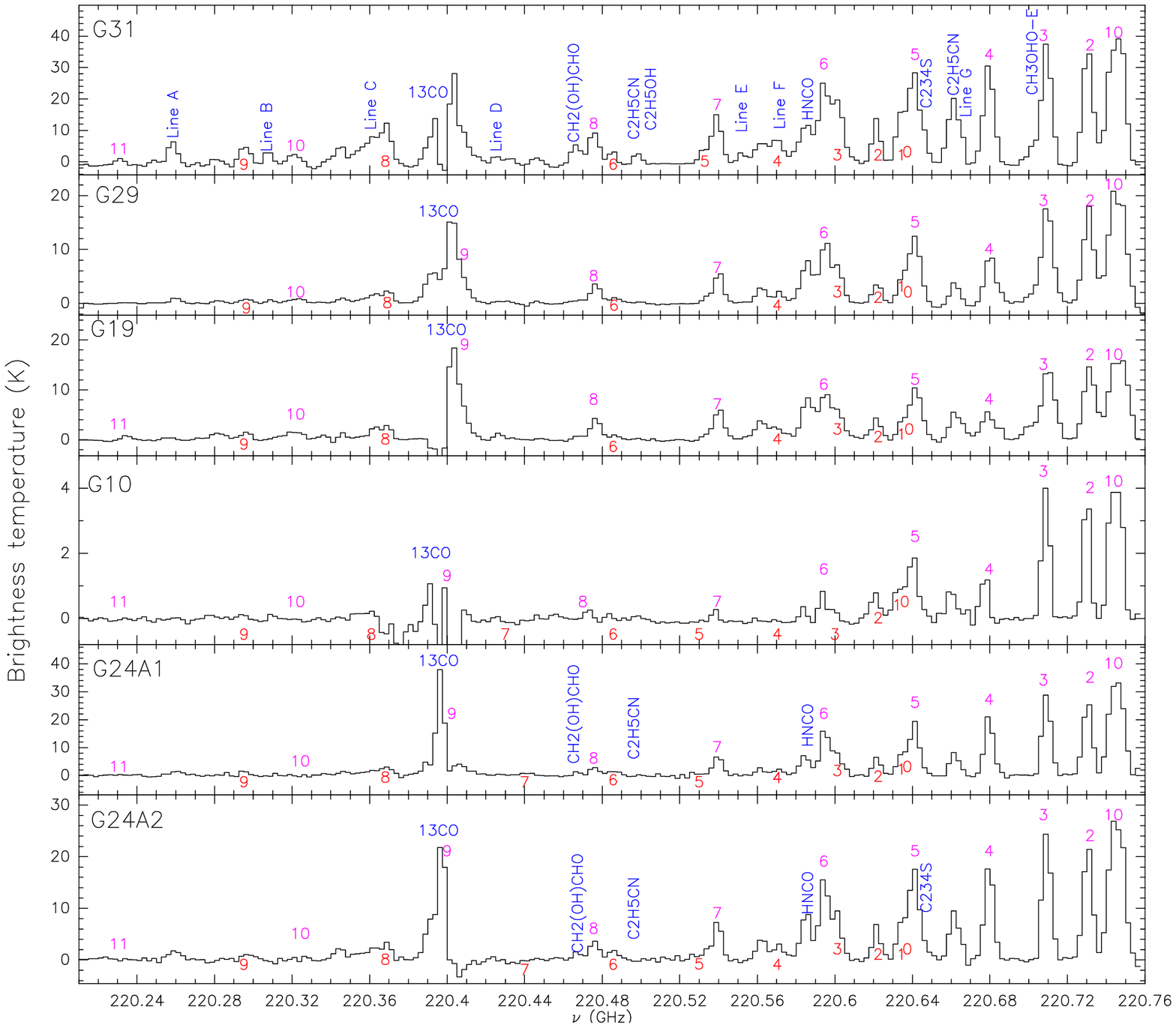}
\end{center}
\caption{Spectra of the hot cores G31, G29, G19, G10, G24A1 and
  G24A2 observed with the PdBI \citep{beltran2005,beltran2011}, integrated over the 3$\sigma$ contour level area. The species labelled are those that have already been identified in \citet{beltran2005} and \citet{beltran2011} as well as the 7 lines in G31 that were found but not identified by \citet{beltran2005}. The numbers indicate the position of the CH$_3$CN~(12$_{\rm K}$--$11_{\rm K}$) K-components   in the upper part (in pink) of each spectra and of the CH$_3$ $^{13}$CN~(12$_{\rm K}$--$11_{\rm K}$) K-components in the lower part (in red).  \label{fig:hotcores}}
\end{figure*}

Figure \ref{fig:hotcores} shows the reduced spectrum for each hot core;
we label the identified lines \citep{beltran2005}  as well as the
seven unidentified lines in G31, A to G in ascending frequency in the
top panel.  As can be seen from the figure, not all lines are present
in all cores.  Lines C, D, F and G were too blended with other lines
to be clearly identified.  Table~\ref{tab:lines} lists the previously
unidentified lines which could be identified (A, B and E), with their
frequencies and observational parameters, and which cores they are
detected in.  These will be discussed in Sect.~\ref{sect:ufolines}.

We also find that several of the molecular lines identified in
G31, G24A1 and G24A2 by \citet{beltran2005}  are also seen in other members of our
sample of hot cores. We briefly discuss these new detections in Sect.~\ref{sect:newdet}.\\

We determine the identity of all of the spectral lines in our sample using a rigorous method to avoid misidentifications. Here we follow the line identification
process as presented in \citet{snyder2005}.  While we refer the reader
to the \citet{snyder2005} paper for a detailed explanation of the
recipe one has to follow in order to identify a spectral line, here we
very briefly summarise the criteria used:

\begin{enumerate}
\item We only considered transitions with a frequency error of $<$50 MHz. 
\item All the expected transitions of a molecule must have frequency
  agreement. In the case of line blending two lines should be at least
  resolved by the Rayleigh criterion. \citet{snyder2005} also remark
    that a more stringent criterion may be used whereby two
    overlapping lines can be considered to be resolved if they are
    least separated at the half-maximum intensity of the weakest line.
\item If other transitions from the same molecule are present in the
  observed frequency range, have line strengths at detectable levels, and there are no mitigating circumstances such as maser activity or line self-absorption as to why they should not be observed, they should be present in the spectra. Their relative intensity must also correspond
  to the predicted one under LTE.
\item Transitions from molecules already observed in similar objects
  and, if not, in the interstellar medium (ISM), should be favoured
  over new molecules not yet detected in space.
\item Excluding transitions with upper energy levels exceeding the highest upper energy level of a transition previously detected in the hot core sample (931 K).

\end{enumerate}

We have used the JPL, CDMS, and Lovas/NIST molecular spectroscopy
databases for line identification \footnote{For references see acknowledgements.}.  We find that there are
uncertainties in both the frequencies of lines in the molecular
databases and the observed frequency of the unidentified astronomical
lines. Any line with a measured laboratory frequency uncertainty
larger than 50\,MHz has been excluded from consideration.
Since the spectral resolution of the observations is 2.5\,MHz, we
searched for lines within the linewidths of the unidentified lines.  The list of potential lines has
been reduced by excluding transitions with an upper energy level over
 931\,K, as typically excitation temperatures of the present hot core
sample do not exceed $\sim$300\,K \citep[e.g.][and references
  therein]{beltran2005,beltran2010} and to date, the highest upper energy level of a transition previously detected in the hot core sample is 931 K. 

\subsection{New detections}
\label{sect:newdet}

In this subsection we identify lines in the G10, G19 and G29 hot cores, which have already been detected in G31, and in most cases already detected in G24A1 and G24A2 as well. Table~\ref{tab:molcon} shows the derived frequency, velocity, integrated intensity, FWHM, peak temperature and rms of the baseline for these newly detected lines in each hot core and the transitional information can be found in Table~\ref{tab:trans}.

\begin{table*}
\caption{The observed frequency, velocity, area of Gaussian fit, FWHM, peak brightness temperature 
and the rms of the baseline for molecules previously detected in \citet{beltran2005} and confirmed in new cores in this work. Spectra are integrated over the 3$\sigma$ contour level.\label{tab:molcon}}
\begin{center}
\scriptsize{
\begin{tabular}{cccccccc}
\hline

Molecule&Frequency$^{\dagger}$& V$_{LSR}$& V$_{peak}$$^{\dagger}$&$\int T$\,d$v$ (Error) & FWHM & T$_\mathrm{B}$ & rms of baseline\\
        &(MHz)&(kms$^{-1}$)&(kms$^{-1}$)&(Kkms$^{-1})$&     (kms$^{-1}$) &(K)&(K)\\
\hline
         &           &               &  G31.41+0.31 && & \\ 
\hline 
CH$_2$OHCHO &220\,466.35 &96.8&93.6 (0.7)&\phantom{1}40.5 \phantom{1}(1.2) &\phantom{1}6.8 (0.1)&\phantom{1}5.6 (0.2) & 0.3\\
HNCO        &220\,585.58 &96.8&96.5 (1.0)&          174.9 (22.5)           &          14.9 (2.4)&          11.1 (0.9) & 0.9\\
C$_2$H$_5$CN&220\,661.45 &96.8&96.3 (0.3)&          203.5 (37.6)           &\phantom{1}9.7 (2.0)&          19.7 (1.2) & 0.3\\
\hline
         &           &                &  G29.96-0.02 & && \\ 
\hline 
CH$_2$OHCHO &220\,467.53 &98.9&93.9 (3.4)&\phantom{1}6.5 \phantom{1}(1.1)  &10.9 (2.1)           &0.6 (0.1) &0.2 \\
HNCO        &220\,585.88 &98.9&98.0 (0.2)&85.6 (18.7)                      &10.5 (2.9)           &7.7 (0.2) &0.3\\
C$_2$H$_5$CN&220\,661.96 &98.9&97.5 (0.4)&29.3 \phantom{1}(9.2)            &\phantom{1}6.9 (2.5) &4.0 (0.1) &0.8\\
\hline
         &           &                &  G19.61-0.23 && & \\ 
\hline 
CH$_2$OHCHO&220\,467.67 &41.6&36.4 (0.2)&\phantom{1}5.8 \phantom{1}(1.4)  &\phantom{1}6.9 (1.9) &0.8 (0.1) &0.2 \\
HNCO&220\,586.00 &41.6&40.5 (0.2)&87.5 (41.4)  &  10.2 (5.5)  &  8.1 (0.3)  & 0.6\\
C$_2$H$_5$CN&220\,662.77 &41.6&39.1 (0.4)&61.0 \phantom{1}(4.2)  &  10.8 (0.9)  &   5.3 (0.5) & 0.9\\

\hline
         &           &       &           G24.78+0.08 A1 && & \\ 
\hline 
CH$_2$OHCHO &220\,467.22 &110.8&106.5 (0.2)&\phantom{1}8.0 \phantom{1}(0.3)  &\phantom{1}4.3 (0.5)  &  1.8 (0.1)  &0.2\\
HNCO        &220\,584.53 &110.8&111.9 (0.2)&67.9 (16.3) &\phantom{1}9.0 (2.5)  &   7.1 (0.3)  &0.8\\
C$_2$H$_5$CN&220\,661.67 &110.8&110.0 (0.2)&68.5 \phantom{1}(9.1)            &\phantom{1}7.8 (1.2)  &  8.2 (0.5)  &1.1\\
\hline
         &           &       &           G24.78+0.08 A2 && & \\ 
\hline
CH$_2$OHCHO &220\,466.94 &110.8&106.8 (0.5)&10.3 \phantom{1}(1.8)  & \phantom{1}5.8 (1.2)  &1.7 (0.1) &0.3 \\
HNCO        &220\,584.65 &110.8&111.8 (0.1)&59.2 (35.6) &  \phantom{1}6.4 (4.1)  &8.7 (1.5) &0.8\\
C$_2$H$_5$CN&220\,661.67 &110.8&110.0 (0.3)&83.0 (13.9) &  \phantom{1}8.4 (1.6)  &9.3 (0.6) &1.0\\
\hline 

\end{tabular}
}
\end{center}
$^{\dagger}$ The small spread in frequency  (0.2 -- 2.4 MHz)  between the rest frequency of the line and the observed frequency for each molecule in each hot core, is due to the difference between the V$_{LSR}$ and the V$_{peak}$ in each hot core. This difference is within the spectral resolution and V$_{peak}$ error of the observations. 
\end{table*}

\begin{table*}
\caption{Transitional information for the molecules previously detected in \citet{beltran2005} and confirmed in new cores in this work. All this transitional information was taken from the JPL spectral line catalog. \label{tab:trans}}
\begin{center}
\footnotesize{
\begin{tabular}{ccccc}
\hline
Molecule& Transition&Frequency &E$_{u}$&S$\mu^2$\\
               &         & (MHz)    &(K)    &(D$^{2}$)\\  
\hline
 
CH$_2$OHCHO&20(2,18)--19(3,17)&220\,463.88&120.05&\phantom{1}89.74\\
HNCO&10(1,9)--9(1,8)&220\,585.20&101.50&\phantom{1}27.83\\
C$_2$H$_5$CN&25(2,24)--24(2,23)&220\,660.92&143.02&367.60\\
\hline 
\end{tabular}
}
\end{center}
\end{table*}

\begin{figure}
\begin{center}
\includegraphics[angle=-90,width=7cm]{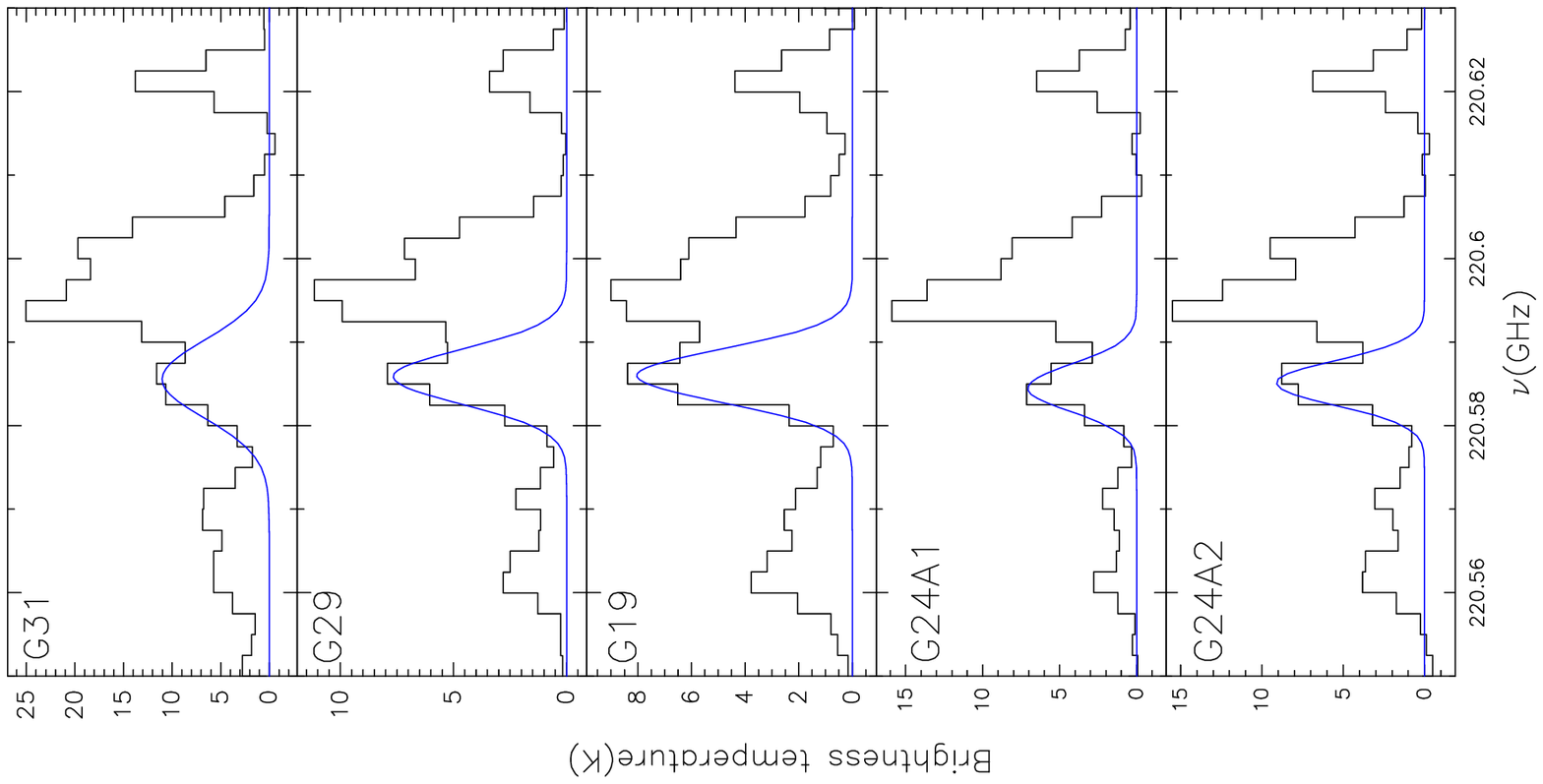}
\end{center}
\caption{HNCO spectra with Gaussian fits, integrated over the 3$\sigma$ contour level area toward G31, G29, G19, G24A1 and G24A2 as seen with the PdBI. This transition was previously detected in G31, G24A1 and G24A2 \citep{beltran2005} and is detected in G19 and G29 in this work.}
\label{fig:HNCO}
\end{figure}

The $10_{1,9}$--$9_{1,8}$ transition of HNCO is detected in a further two hot cores, G19 and G29 (see Fig.~\ref{fig:HNCO}), although it is blended with one transition of methyl cyanide in all cores. 

\begin{figure}
\begin{center}
\includegraphics[angle=-90,width=7cm]{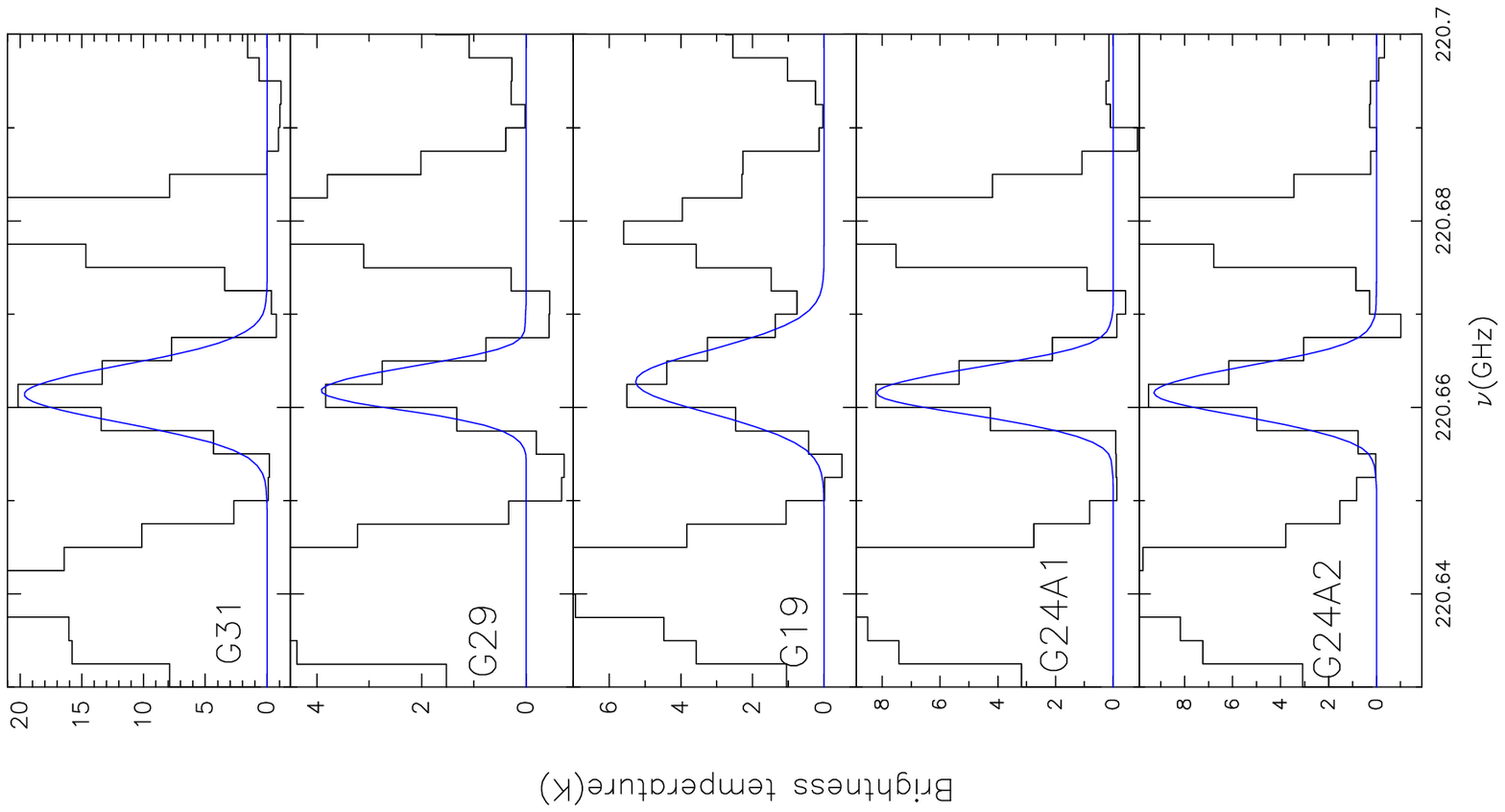}
\end{center}
\caption{C$_2$H$_5$CN spectra with Gaussian fits, integrated over the 3$\sigma$ contour level area toward G31, G29, G19, G24A1 and G24A2 as seen with the PdBI. This transition was previously detected in G31, G24A1 and G24A2 in \citet{beltran2005} and is detected in G19 and G29 in this work.\label{fig:C2H5CN}}
\end{figure}

The $25_{2,24}$--$24_{2,23}$ transition of C$_2$H$_5$CN was also detected
in G31, G24A1 and G24A2 in \citet{beltran2005}.  We have found it in a further two hot
cores: G19 and G29 (Fig.~\ref{fig:C2H5CN}). This line is very bright in G31 at 19.7\,K but is as
weak as 4.0\,K in G29. The FWHM is similar in all the hot cores,
ranging from 6.9\,kms$^{-1}$ in G29 to 10.8\,\,kms$^{-1}$ in G19. It is not detected in G10. Both of these lines are brightest in G31.

\subsubsection{Glycolaldehyde (CH$_2$(OH)CHO)}

The complex organic molecule glycolaldehyde (CH$_2$OHCHO), which is an isomer of both methyl formate (HCOOCH$_{3}$) 
and acetic acid (CH$_3$COOH), is the simplest of the monosaccharide sugars. This important organic molecule 
was first tentatively detected at 1.4~mm with the PdBI outside the Galactic centre, where it was observed in Sgr B2(N) \citep{hollis2000} towards three massive hot-cores (G31, G24A1, and G24A2) by \citet{beltran2005}.
Later on, \citet{beltran2009} confirmed the detection by observing two additional Glycolaldehyde transitions
towards G31 at 2.1 and 2.9~mm with the PdBI.
Recently, \citet{jorgensen2012} have detected 13 transitions towards the Class 0 IRAS 16293--2422 object at 
0.4 and 1.4~mm, using the ALMA array towards the hot-corino surrounding the Solar-type protostar. 

\begin{figure}
\begin{center}
\includegraphics[angle=-90,width=8cm]{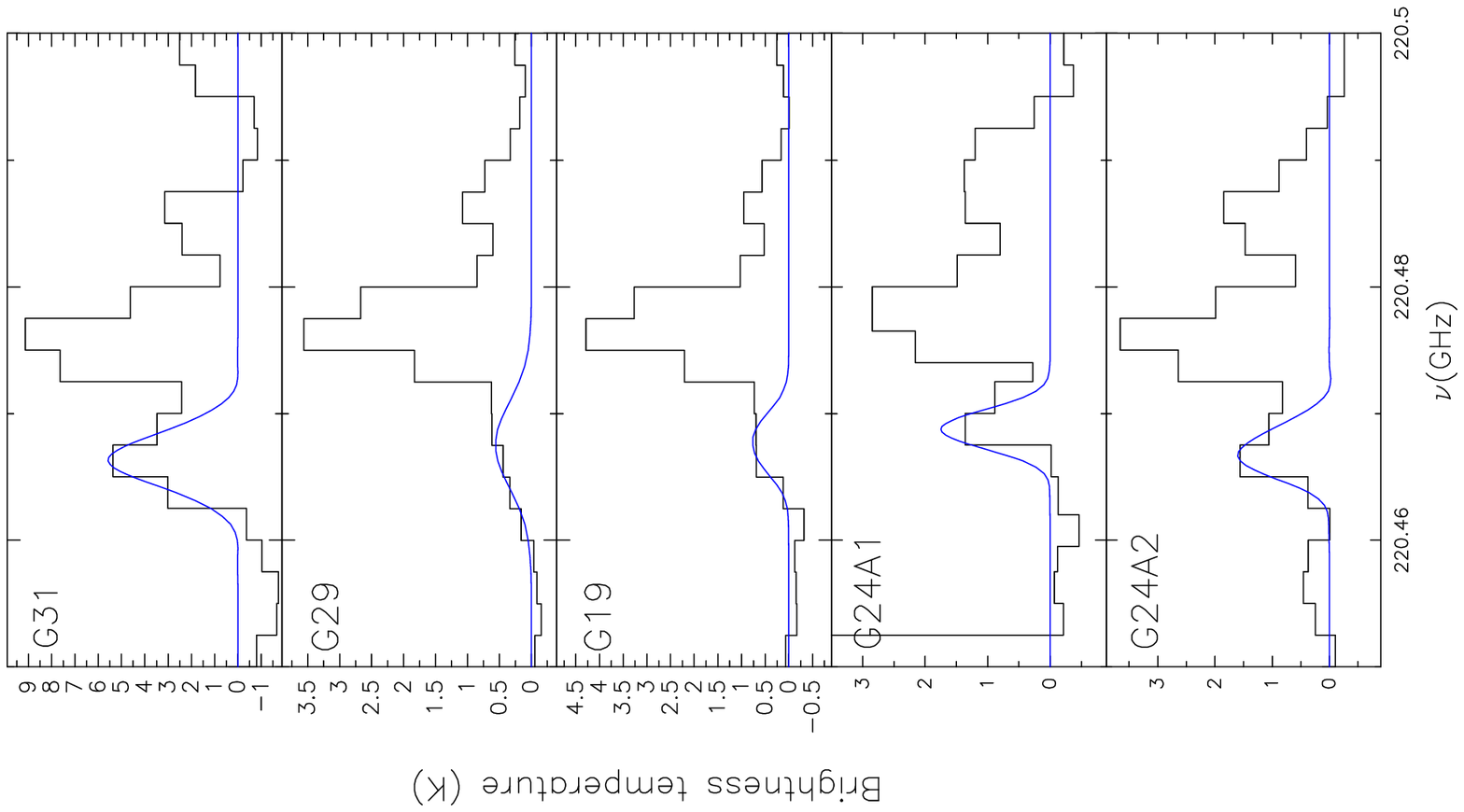}
\end{center}
\caption{CH$_2$(OH)CHO spectra with Gaussian fits, integrated over the 3$\sigma$ contour level area toward G31, G29, G19, G24A1 and G24A2 as seen with the PdBI. This transition was previously detected in G31, G24A1 and G24A2 in \citet{beltran2005} and is detected in G19 and G29 in this work. In G29 it is only detected at a $3\sigma$ level.  More transitions of glycolaldehyde are needed to confirm its presence in G29 and G19, and this line may be contaminated by the 46$_{20,26}$-46$_{19,27}$ EE and 11$_{11, 1, 1}$ -- 10$_{10, 0, 1}$AE   transitions of acetone.\label{fig:glycol}}
\end{figure}

 In this present work, we report a new detection of a line at $\sim$220 466 MHz in two hot cores (Fig.~\ref{fig:glycol}), G29 (3$\sigma$ detection) and G19 (5$\sigma$ detection), which we postulate to be the $20_{2,18}$--$19_{3,17}$ transition  of glycolaldehyde, as was detected by \citet{beltran2005} in G31, G24A1 and G24A2. We observe a shift in the velocity of  3-4 km/s in this transition compared to the V$_{LSR}$ of each hot core. This is consistent with the glycolaldehyde observations of both \citet{beltran2009}, \citet{halfen2006}, and \citet{hollis2000}. Whilst further transitions of this molecule are needed to confirm its presence within G29 and G19, if confirmed, it would suggest that  glycolaldehyde is a common hot-core tracer. In both hot-cores this line is blended with two methyl cyanide lines, as it is in G31, G24A1 and G24A2. We also acknowledge that the 46$_{20,26}$--46$_{19,27}$ EE (E$_u$ = 816 K, S$\mu$$^2$ = 2843 D$^2$) and 11$_{11, 1, 1}$ -- 10$_{10, 0, 1}$AE (E$_u$ = 63 K, S$\mu$$^2$ = 519 D$^2$) transitions of acetone may be contributing to the emission seen at this frequency. This is consistent with observations by \citet{fuente2014}, where they detect the $20_{2,18}$--$19_{3,17}$ transition  of glycolaldehyde at 220 466 MHz, blended with acetone. In Section \ref{sec:specmod} we model both glycolaldehyde and acetone emission to explore this possibility. For the purposes of a chemical comparison between the hot cores in our sample, we have assumed that the line at 220 466 MHz is glycolaldehyde in the rest of the paper. 

\begin{figure*}
\begin{center}
\includegraphics[angle=-90,width=16cm]{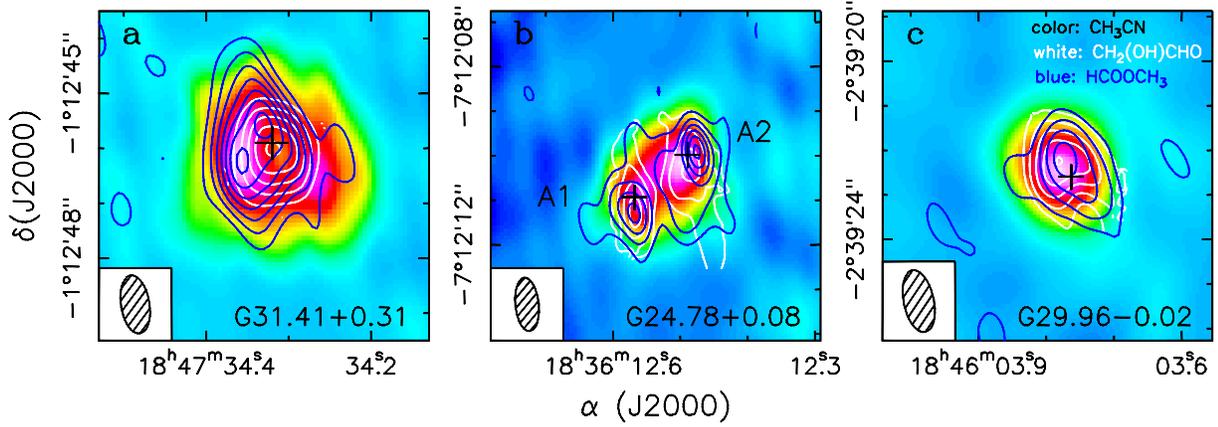}
\end{center}
\caption{Spectral line map of the 18$_{8,10}$-17$_{8,9}$E transition of methyl formate
  (blue contours), the  20$_{2,18}$--19$_{3,17}$ transition of glycolaldehyde (white contours) and the averaged emission of the K=0, 1, 2 (12-11) transitions of methyl cyanide emission (colour scale)  in G31, G29, G24A1 and G31. \label{fig:formatemap} For the methyl formate in G31, the channels averaged were 91.7 -- 101.9 kms$^{-1}$, with contour levels of 0.04 -- 0.28 Jybeam$^{-1}$, in steps of 0.04 Jybeam$^{-1}$. In G24 the channels averaged for methyl formate were 102.4 -- 112.6 kms$^{-1}$, with contour levels of 0.02 -- 0.12 Jybeam$^{-1}$, in steps of 0.02 Jybeam$^{-1}$. In G29 the channels averaged for methyl formate were 91.1 -- 101.3 kms$^{-1}$, with  contour levels of 0.015 -- 0.06 Jybeam$^{-1}$, in steps of 0.015 Jybeam$^{-1}$. For glycolaldehyde in G31,  the channels averaged were 91.9 -- 95.3 kms$^{-1}$ with contour levels of  0.10 -- 0.46 Jybeam$^{-1}$, in steps of 0.12 Jybeam$^{-1}$. In G24 the channels averaged for glycolaldehyde were 104.8 -- 108.2 kms$^{-1}$, with contour levels of 0.04 -- 0.12 Jybeam$^{-1}$, in steps of 0.04 Jybeam$^{-1}$. In G29 the channels averaged for glycolaldehyde were 92.2 -- 95.6 kms$^{-1}$, with contour levels of 0.016 -- 0.08 Jybeam$^{-1}$, in steps of 0.016 Jybeam$^{-1}$. For methyl cyanide the contour levels in G31 are 0.1--0.94 Jybeam$^{-1}$  in steps of 0.12, in G24 they are 0.1 -- 1.0 Jybeam$^{-1}$ in steps of 0.18 Jybeam$^{-1}$, in G29 they are 0.09 -- 1.14 Jybeam$^{-1}$ in steps of 0.15 Jybeam$^{-1}$. }
\end{figure*}

Figure~\ref{fig:formatemap} shows a map of the $18_{8,10}$--$17_{8,9}$E
methyl formate emission (blue contours; line A, see Section \ref{sect:ufolines}), the
$20_{2,18}$--$19_{3,17}$ glycolaldehyde emission (white contours) and 
the averaged emission of the K=0, 1, 2 (12--11) transitions of methyl cyanide emission (colour scale)  in G31,
G24A1, G24A2 and G29 (i.e. the four hot-cores for which the three molecular species have been detected). Methyl cyanide is a typical hot core tracer.
For G24A1, G24A2, and G29 the three species seem to be tracing the same material and peak at roughly the
same position towards the dust emission peak. On the other hand, in G31 glycolaldehyde peaks towards the
centre of the core where the continuum source(s) is(are) embedded, whilst 
methyl formate and methyl cyanide show a different morphology.
In particular, as reported by \citet{beltran2005} the methyl cyanide traces a 
toroidal structure with the strongest emission eastwards  of the millimetre continuum emission peak, which is located towards the central dip.
On the other hand, methyl formate does not show a toroidal morphology but peaks towards the
eastern side of the core, at a position barely coincident with that of the methyl cyanide.
\citep{beltran2005} explained these apparent toroidal morphology as caused by the high 
optical depth and the existence of a temperature gradient in the core.
In this scenario, glycolaldehyde appears to be less affected by excitation conditions,
and a better tracer of the inner conditions of the hot-core closer to the embedded protostar(s). Methyl formate emission extends to 0.13\, pc (3\farcs5) in G31 where the glycolaldehyde extends to 0.08\, pc (2\arcsec). In G24, both methyl formate emission and glycolaldehyde emission extend to 0.11\, pc (3\arcsec). In G29, both methyl formate emission and glycolaldehyde emission extend to 0.05\, pc (3\arcsec). 

\subsection{New identifications}
\label{sect:ufolines}
\begin{figure*}
\begin{center}
\includegraphics[width=9cm, angle=-90]{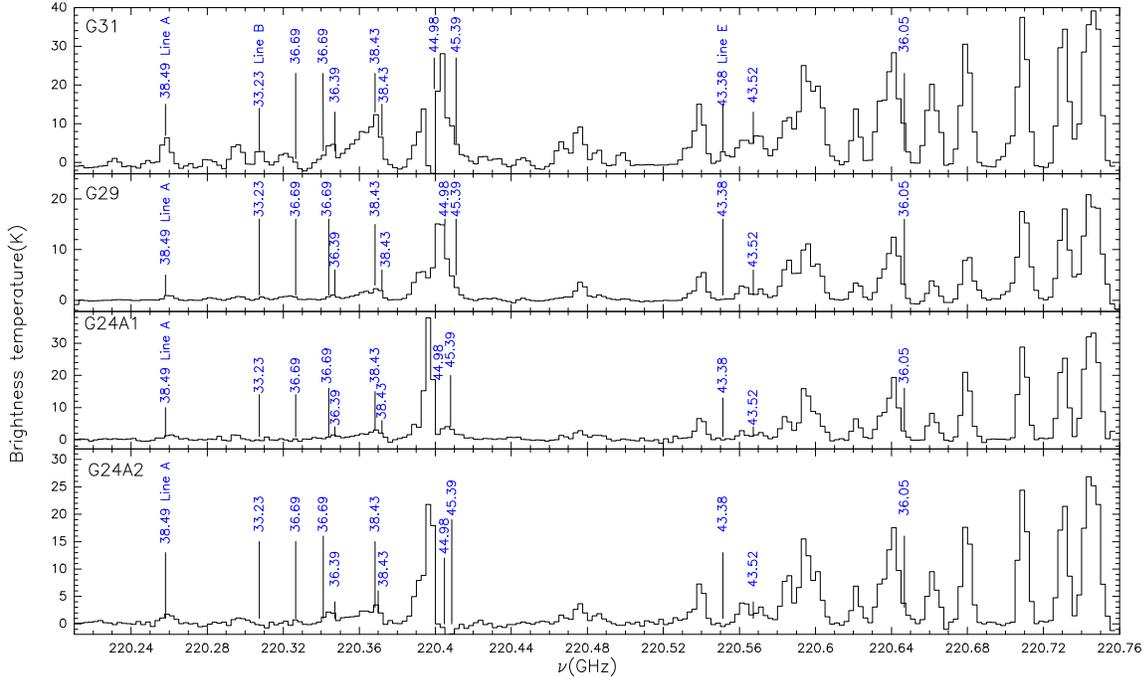}
\end{center}
\caption{The expected bright emission of methyl formate transitions, that occur in the frequency range of our observations, plotted on the spectra of hot cores G31, G29, G24A1 and G24A2. The numbers on the plot represent the intensity (S$\mu^{2}$) of methyl formate transitions as determined by laboratory studies, in units of Debye$^2$. \label{fig:lineAmethylformate} }
\end{figure*}
\begin{figure}
\begin{center}
\includegraphics[angle=-90,width=8cm]{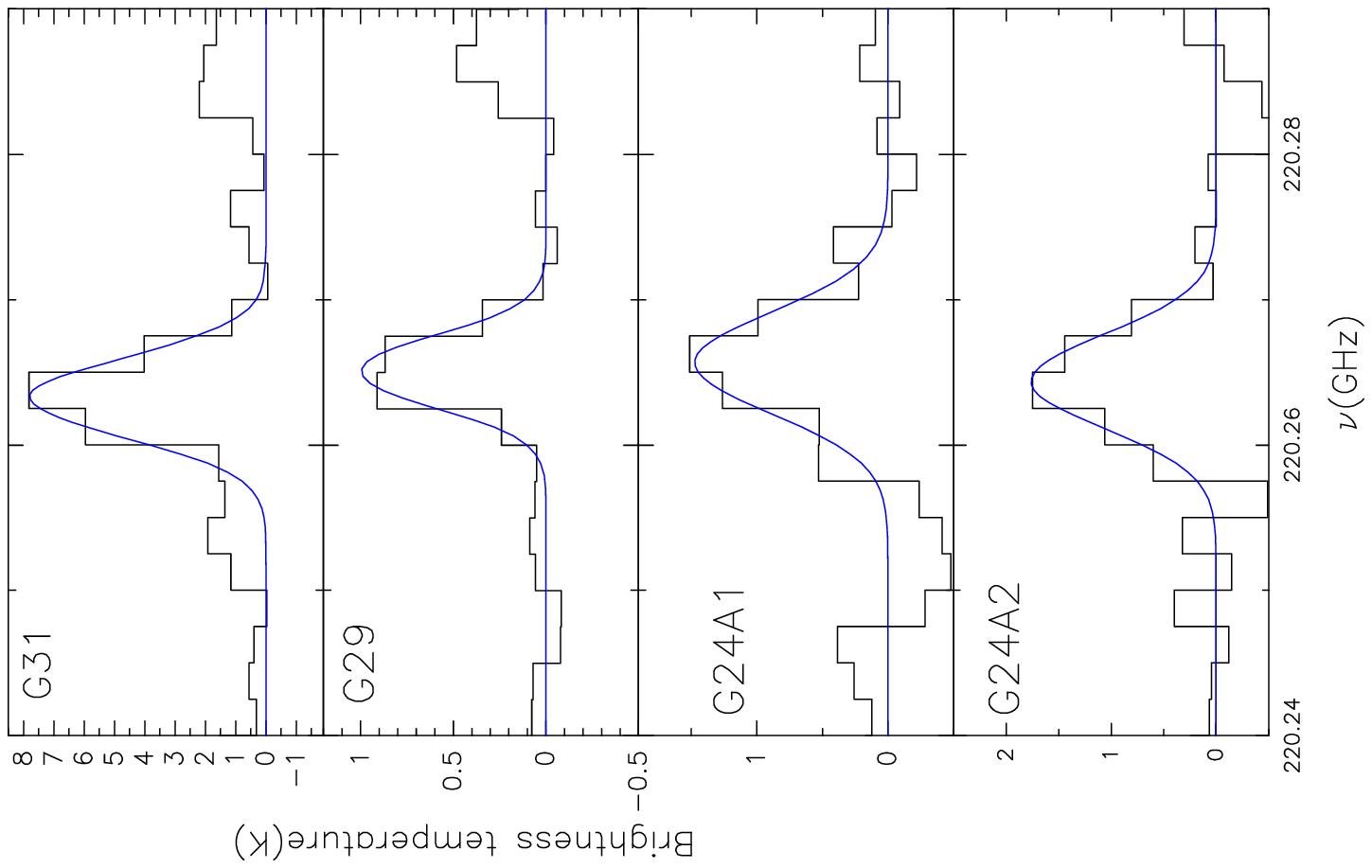}
\end{center}
\caption{Spectra of Line A with Gaussian fits, integrated over the 3$\sigma$ contour level area toward G31, G29, G24A1 and G24A2 as seen with the PdBI. This transition was previously detected in G31 but was not identified, in \citet{beltran2005}, and is also detected in G24A1, G24A2 and G29 in this work. \label{fig:lineA} }

\end{figure}
We now discuss some of the lines that were previously unidentified
(see Fig.~\ref{fig:hotcores}), namely A, B and E , for each source,
and we compare the frequency, area of the Gaussian fitted, FWHM and
peak brightness temperature in each hot core for each of the lines
(see Table~\ref{tab:lines}).   It is clear from Fig.~\ref{fig:hotcores}
that the spectra do vary among hot cores and that G31, the brightest
hot core with $L_\mathrm{bol}>10^{4}L_{\odot}$
\citep{beltran2005} is the most chemically-rich hot core in our
sample. Lines C, D, F and G labelled in this figure could not be identified.

Line A is seen in four of the hot cores in our sample: G31, G29, G24A1
and G24A2 (Fig.~\ref{fig:lineA}). We have fitted a Gaussian profile to
this line in each of the hot cores to determine their rest frequency,
integrated intensity, FWHM, and brightness temperature (see
Table~\ref{tab:lines}). There is only a small spread in the observed
rest frequency of this line from 220\,258.240\,MHz to 220\,260.657\,MHz,
suggesting it is the same line in each hot core. Line A is brightest
in G31, as expected. The FWHM range from 7.6\,kms$^{-1}$ in G29 to 11.1\,kms$^{-1}$
in G24A1.

\begin{table*}
\caption{The observed frequency, velocity, area of Gaussian fit, FWHM, peak brightness temperature 
and the rms of the baseline for lines A, B and E in each of the hot cores in our sample. Spectra are averaged over the 3$\sigma$ contour level.\label{tab:lines} }
\begin{center}
\footnotesize{
\begin{tabular}{cccccccc}
\hline 
Object&Frequency$^{\dagger}$ &V$_{LSR}$ & V$_{peak}$&$\int $T\,d$v$ (Error) &FWHM &T$_\mathrm{B,peak}$ & rms of baseline \\
      &   (MHz) &(kms$^{-1}$)  &(kms$^{-1}$)&(K.kms$^{-1}$)     &  (kms$^{-1}$)& (K)&(K)\\
\hline
\multicolumn{6}{|c|}{Line A} \\
\hline
G31&220\,258.26&96.8&96.8&72.0 (1.8)& 8.2 (0.3)& 7.8 (0.2)&0.3\\
G29&220\,260.05&98.9&96.2&\phantom{1}8.3 (0.3)& 7.6 (0.3)  & 1.02 (0.03)&0.2\\
G19&&&&No detection rms = 0.2&&&\\
G10&&&&No detection rms = 0.1&&&\\
G24A1&220\,260.65&110.8&107.5 &17.4 (2.9)& 11.1 (2.2)  & 1.5 (0.2)&0.2\\
G24A2&220\,259.28&110.8&109.2&19.4 (2.1) &10.3 (1.2) &1.8 (0.2)&0.3\\

\hline
\multicolumn{6}{|c|}{Line B} \\
\hline

G31&220\,307.69&96.8& 96.6&37.4 (2.0)& 7.1 (0.5) & 5.0 (0.3)&0.3\\
G29&&&&No detection rms = 0.2&&&\\
G19&&&&No detection rms = 0.2&&&\\
G24A1&&&&No detection rms = 0.2&&&\\
G24A2&&&&No detection rms = 0.3&&&\\
G10&&&&No detection rms = 0.1&&&\\
\hline
\multicolumn{6}{|c|}{Line  E} \\
\hline
G31&220\,552.70&96.8&95.1&16.0 (4.5)&5.0 (2.8)&2.56 (0.9)&0.9\\
G29&&&&No detection rms = 0.3&&&\\
G19&&&& No detection rms = 0.6&&&\\
G10&&&&No detection rms = 0.2&&&\\
G24A1&&&&No detection rms = 0.8&&&\\
G2A2&&&&No detection rms = 0.8&&&\\
\hline
\end{tabular}
}
\end{center}
 $^{\dagger}$ The small spread in frequency (0.2 -- 2 MHz) between the rest frequency of the line and the observed frequency for each molecule in each hot core, is due to the difference between the V$_{LSR}$ and the V$_{peak}$ in each hot core. This difference is within the spectral resolution and V$_{peak}$ error of the observations.

\end{table*}

\begin{table*}
\caption{The observed lines transitions, E$_u$, S$\mu^2$, and line list used for lines A, B and E.\label{tab:untran} }
\begin{center}
\footnotesize{
\begin{tabular}{ccccccc}
\hline
Line &Molecule& Transition&Frequency &E$_{u}$&S$\mu^2$&Line list\\
       &        &         & (MHz)    &(K)    &D$^{2}$\\  
\hline
 
Line A&HCOOCH$_3$ $v$=1& 18(8,10) -- 17(8,9)E&220\,258.09&331&38.5&JPL\\
Line B&HCOOCH$_3$ $v$=1& 18(10,9) -- 17(10,8)E&220\,307.38&354&33.2&JPL\\
Line E&H$^{13}$COOCH$_3$& 18(6,12) -- 17(6,11) ++$v$=1-1&220\,551.30&313&43.9&TopModel\\
\hline \\
\end{tabular}
}
\end{center}
\end{table*}

The potential identities of line A consist of three lines of
methyl formate (HCOOCH$_3$): 18$_{8,10}$-17$_{8,9}$E,
24$_{2,23}$-24${1,24}$E, and 24$_{2,23}$-24$_{0,24}$E; one line of
methylene amidogen (H$_2$CN): 3$_{2,1}$-2$_{2,0}$,F=11/2-9/2,
(n=0-0)*; and one line of vinyl alcohol (C$_2$H$_3$OH):
11$_{2,10}$-11$_{0,11}$.  Methylene amidogen has previously been found
in the cold core TMC-1 \citep{ohishi1994} and methyl formate and vinyl
alcohol have previously been detected in hot cores
\citep{fontani2007,turner2001}. Using criterion (iii) from Section \ref{sec:molcon} we can rule out methylene amidogen as there are several expected transitions in the frequency range of
our observations that are not seen.  Vinyl alcohol can also be ruled
out because the $11_{2,10}$--$11_{0,11}$ transition has a low line intensity (S$\mu^2$ =
0.03), which leads to a column density of
2.7$\times$10$^{20}$\,cm$^{-2}$, six orders of magnitude higher than
the vinyl alcohol column density in Sagittarius B2(N)
\citep{turner2001} and much larger than those of more
commonly-observed molecules. We therefore identify line A with the
18$_{8,10}$--17$_{8,9}$E transition of methyl formate, but the
24$_{2,23}$--24$_{1,24}$E, and 24$_{2,23}$--24$_{0,24}$E transitions also
contribute to the line A emission. The $18_{8,10}$--$17_{8,9}$E transition of methyl formate is a high energy transition therefore it was not identified previously.  

One notes however that, while all the expected transitions of methyl
formate in our frequency range are detected in G31, this is not the
case for the other hot cores (see Fig.~\ref{fig:lineAmethylformate}). For
example the 220\,551.31\,MHz 18$_{6,12}$--17$_{6,11}$, vt=1
transition of methyl formate falls in a region of no detectable
emission in the hot cores G29, G24A1 and G24A2.  This may be an
indication of temperature differences among hot cores as the intensity
of methyl formate transitions is significantly temperature dependent.
\\

\begin{figure}
\begin{center}
\includegraphics[width=4.5cm, angle=-90]{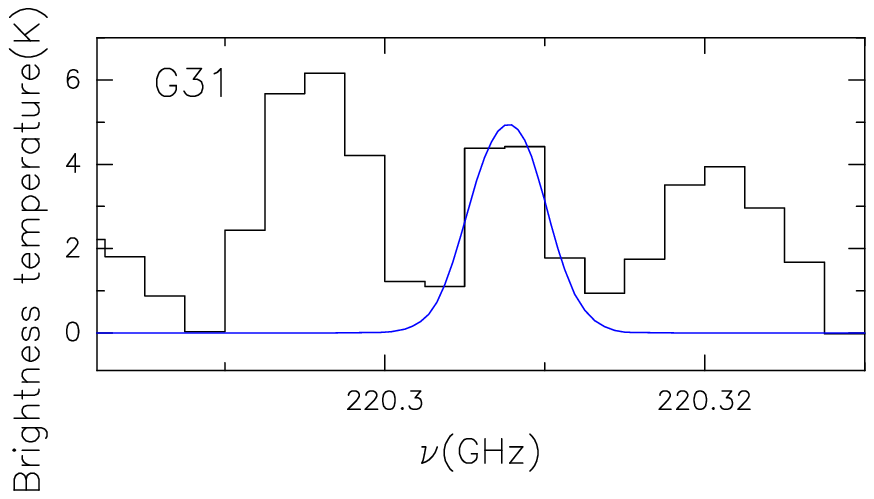}
\end{center}
\caption{Spectra of Line B with a Gaussian fit, integrated over the 3$\sigma$ contour level area toward G31 as seen with the PdBI. This transition was previously detected in G31 but was not identified in \citet{beltran2005}. \label{fig:lineB} }
\end{figure}

Line B is only detected in G31 (see Fig.~\ref{fig:lineB}). Its
potential line identities are three transitions of methylene amidogen
(H$_2$CN) and one line of methyl formate
(18$_{10,9}$--17$_{10,8}$E). Methylene amidogen can, again, be ruled out according to the (iii) criterion of Section \ref{sec:molcon},
so the methyl formate 18$_{10,9}$--17$_{10,8}$E transition is the best
candidate for line B.\\

\begin{figure}
\begin{center}
\includegraphics[width=4cm, angle=-90]{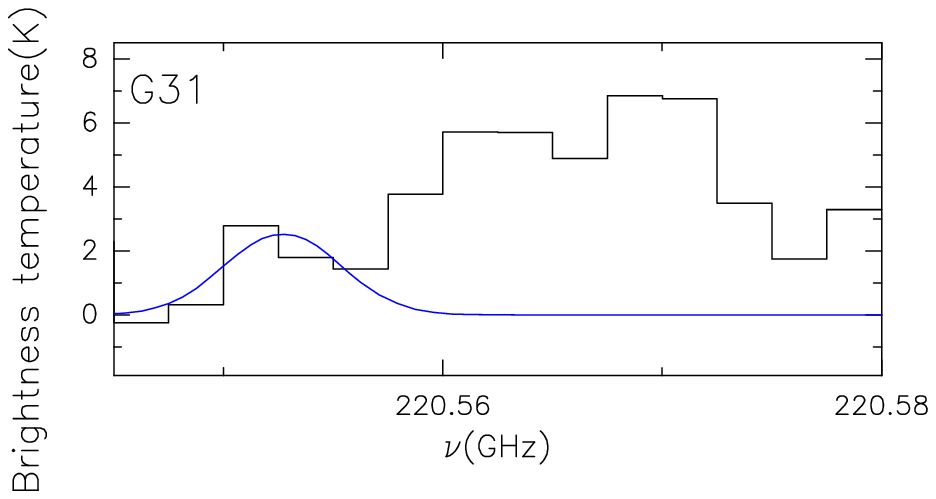}
\end{center}
\caption{Spectra of Line E with a Gaussian fit, integrated over the 3$\sigma$ contour level area toward G31 as seen with the PdBI. This transition was previously detected in G31 but was not identified in \citet{beltran2005}. \label{fig:lineE}}

\end{figure}

Line E is only detected in G31 (see Fig.~\ref{fig:lineE}). There are
two potential line identities for line E, one line of a carbon-13
isotopologue of methyl formate (H$^{13}$COOCH$_3$),
18$_{6,12}$--17$_{6,11}$, vt=1 and one line of propanal
(CH$_3$CH$_2$CHO), 9$_{4,5}$--8$_{2,6}$. There is only one bright line
of propanal in the frequency range of these observations. This line
falls in an area of little emission in the spectrum of G31 thereby ruling out propanal
as a candidate. The 18$_{6,12}$--17$_{6,11}$, vt=1 transition of
methyl formate is therefore the best candidate for line E. The transitional information for these newly identified lines can be found in Table \ref{tab:untran}.\\

Clearly G31 is the most chemically rich object and in fact, of the three unidentified lines, B and E are only present in this source.  In the next sections we derive column densities assuming Local Thermodynamic Equilibrium (LTE) and analyse our results through the use of a chemical model.

\section{Spectral line analysis}
\subsection{Estimates of excitation temperature}
\label{sec:mf}
\begin{figure}
\begin{center}

\includegraphics[width=8cm]{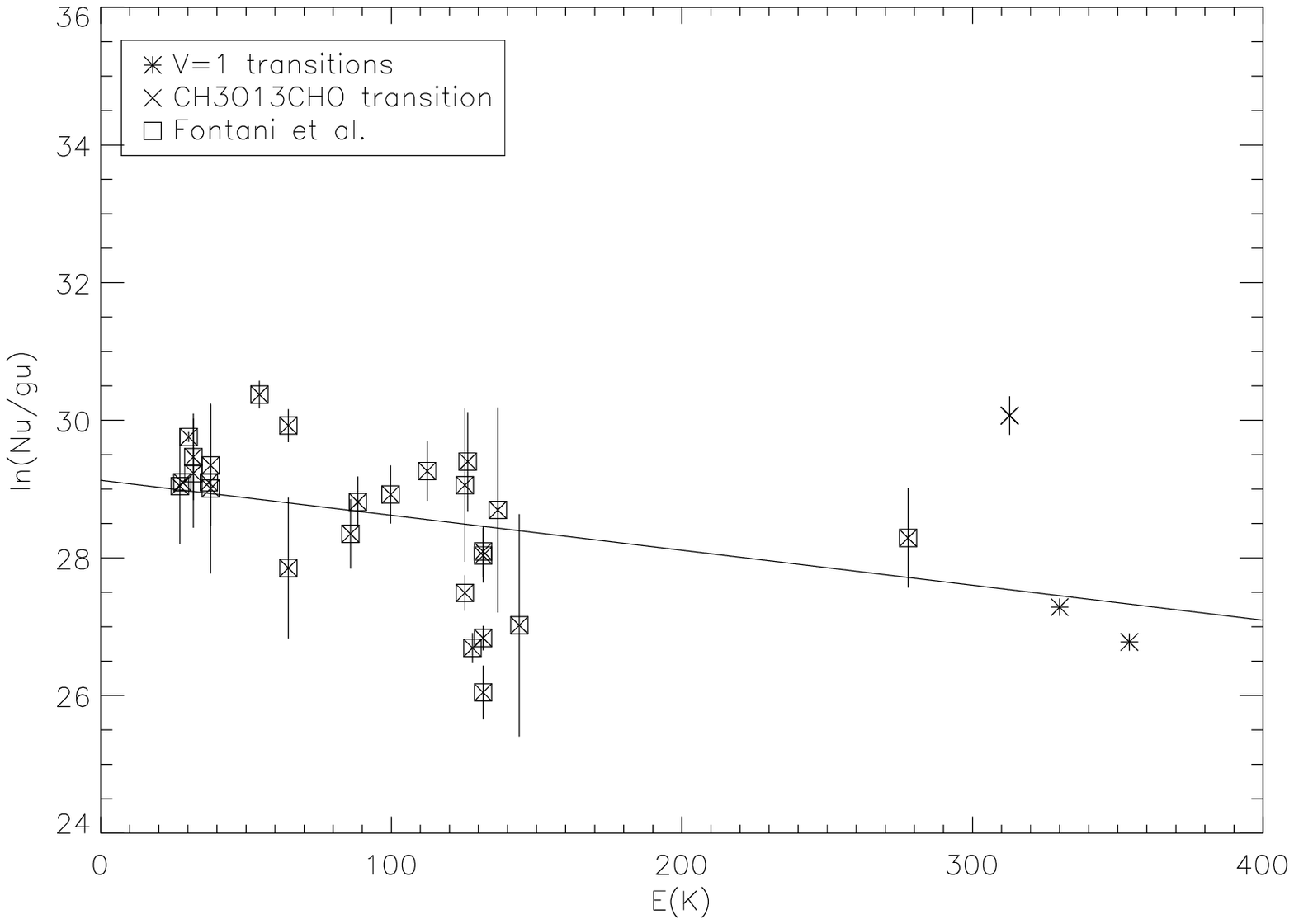}
\end{center}
\caption{A rotation diagram of methyl formate transitions in G31, extending
  the work of \citet{fontani2007} to higher excitation
  temperatures. A $^{12}$C/$^{13}$C ratio of 41 was calculated according to \citet{wilson1994} using the Galactic Coordinates of G31 and a source size of 3\farcs5  was assumed based on the observed distribution seen in Fig.~\ref{fig:formatemap}. Errors for the $v$=1 excited transitions are so small ($<$0.02) that they do not appear on this diagram.}\label{fig:rotdiaformate}
\end{figure}

\begin{table*}
\caption{Estimated rotational temperatures and column density of
  methyl formate calculated using the rotational temperature and using a source size measured in this work.}\label{tab:methylformate}
\begin{center}
\footnotesize{
\begin{tabular}{ccccccc}
\hline
Object&Source size (\arcsec)& T$_\mathrm{rot}$ (K) & Column density (cm$^{-2}$)\\
\hline
G31&3.5&169$\pm39$&4$\times$10$^{17}$\\
\hline
\end{tabular}
}
\end{center}
\end{table*}

Methyl formate in G31 has been extensively studied and, aside from 
three newly-identified transitions, 26 transitions have been
previously detected \citep{fontani2007} using the IRAM-30m. We therefore use the 29 lines of methyl formate in G31 to derive column densities and temperature estimates using a rotational diagram.  We have accounted for the beam dilution of the IRAM-30m data using a source size of 3\farcs5  as measured in Fig.~\ref{fig:formatemap}. For the 3 newly-identified transitions we assume their emission fills the beam. The results can be seen in
Fig.~\ref{fig:rotdiaformate} and Table~\ref{tab:methylformate}. The linear fits to the data in Fig.~\ref{fig:rotdiaformate} have
    been calculated using a regression line, and the derived rotation
    temperatures agree very well with those of
    \citet{fontani2007}. The spread of data points from this regression line are within the error bars. The derived column density
    (Table~\ref{tab:methylformate}) is  in good agreement with the results of \citet[][5.3 $\times$ 10$^{17}$\,cm$^{-2}$]{fontani2007} and \citet[][1.7 $\times$ 10$^{17}$\,cm$^{-2}$]{isokoski2013} when assuming a source size of 3\farcs5.

From the rotational diagram we find that, surprisingly, a single gas
component with a temperature of $\sim$169\,K can fit all 29
transitions. Note that line A and line B are vibrationally excited
transitions of methyl formate and we would have expected them to trace a hotter region of gas. 
\subsection{Column density estimates}

Table~\ref{tab:cd} lists calculated column densities for all
identified molecules across our sample, derived using Equation \ref{eq:cd} below, by assuming LTE at a
constant temperature of 300\,K, 225\,K and 150\,K, and optically thin emission (see Section \ref{sec:mf} for a detailed discussion of the temperature of our hot cores).

\begin{equation}
N_{u}/g_{u}=\frac{N_{tot}}{Q(T_{rot})}e^{-E_{u}/T_{rot}}=\left(\frac{8\pi\nu^{2}k\int Tdv}{hc^{3}Ag_{u}}\right)
\label{eq:cd}
\end{equation}

where g$_{u}$ is the statistical weight of the level {\it u}, N$_{tot}$ is the total column density of the molecule, Q(T$_{rot}$) is the rotational partition function, E$_{u}$ is the energy of the upper energy level, {\it k} is the Boltzmann constant, {\it$\nu$} is the frequency of the line transition, {\it A} is the Einstein coefficient of the transition, $\int Tdv$ is the integrated line intensity. All column densities have been corrected for beam dilution by dividing the integrated line intensity by the beam dilution factor (Equation \ref{eq:bd}).

\begin{equation}
\eta_{BD}=\frac{\theta_S^2}{\theta_S^2+\theta_B^2}
\label{eq:bd}
\end{equation} 

where $\theta_S$ is the source size and $\theta_B$ is the beam size.

For the IRAM-30m observations of G31 we use a beam size ranging from 10\arcsec -- 24\farcs4. For the PdBI observations we assume the source fills the beam.

\begin{table*}
\caption{Column densities (cm$^{-2}$) of organic molecules in our sample, assuming
  LTE at 300\,K, 225\,K, and 150\,K. \label{tab:cd} }
\begin{center}
\begin{tabular}{ccccccc}
\hline
 & & &300\,K&&&\\
\hline
Object  &         HNCO                           & C$_2$H$_5$CN  & CH$_3$CN$^{a}$                      & CH$_2$(OH)CHO                               &HCOOCH$_{3}$ \\ 
\hline
G31 &  3$\times$10$^{16}$ & 1$\times$10$^{16}$ &  5$\times$10$^{16}$ & 1$\times$10$^{17}$$^{\dagger\dagger}$& 1$\times$10$^{18}$$^{\dagger\dagger\dagger}$ \\
G29 &  2$\times$10$^{16}$ & 2$\times$10$^{15}$ & 2$\times$10$^{16}$  & 5$\times$10$^{15}$  \phantom{1}                           & 2$\times$10$^{16}$\phantom{11}  \\
G19 &  1$\times$10$^{16}$ & 3$\times$10$^{15}$ &1$\times$10$^{16}$ & 3$\times$10$^{15}$     \phantom{1}                         & \ldots \phantom{11}  \\
G24A1 & 1$\times$10$^{16}$ & 4$\times$10$^{15}$  & 4$	\times$10$^{16}$ &5$\times$10$^{15}$            \phantom{1}                & 3$\times$10$^{16}$\phantom{11}  \\
G24A2 & 2$\times$10$^{16}$ & 5$\times$10$^{15}$ & 3$\times$10$^{16}$   &6$\times$10$^{15}$   \phantom{1}                          & 4$\times$10$^{16}$\phantom{11}   \\
\hline
 & & &225\,K&&\\
\hline
G31 &   2$\times$10$^{16}$ & 7$\times$10$^{15}$ &  3$\times$10$^{16}$ & 5$\times$10$^{16}$$^{\dagger\dagger}$           & 8$\times$10$^{17}$$^{\dagger\dagger\dagger}$ \\
G29 &  1$\times$10$^{16}$ & 1$\times$10$^{15}$ & 1$\times$10$^{16}$   & 3$\times$10$^{15}$   \phantom{1}                       & 1$\times$10$^{16}$ \phantom{11}  \\
G19 &  1$\times$10$^{16}$ & 2$\times$10$^{15}$ & 8$\times$10$^{15}$   & 2$\times$10$^{15}$    \phantom{1}                       & \ldots \phantom{11} \\
G24A1 &   7$\times$10$^{15}$ & 3$\times$10$^{15}$  & 3$\times$10$^{16}$  &5$\times$10$^{15}$     \phantom{1}                       & 2$\times$10$^{16}$\phantom{11}  \\
G24A2 &   8$\times$10$^{15}$ & 3$\times$10$^{15}$ & 2$\times$10$^{16}$  &6$\times$10$^{15}$       \phantom{1}                      & 2$\times$10$^{16}$\phantom{11}   \\
\hline
 & & &150\,K&&\\
\hline
G31 &   1$\times$10$^{16}$ & 4$\times$10$^{15}$ &  2$\times$10$^{16}$ & 2$\times$10$^{16}$$^{\dagger\dagger}$            & 4$\times$10$^{17}$$^{\dagger\dagger\dagger}$ \\
G29 &    7$\times$10$^{15}$ & 6$\times$10$^{14}$ & 5$\times$10$^{15}$ & 1$\times$10$^{15}$   \phantom{1}                         & 5$\times$10$^{15}$\phantom{11}   \\
G19 &  5$\times$10$^{15}$  & 1$\times$10$^{15}$ & 4$\times$10$^{15}$ & 8$\times$10$^{14}$   \phantom{1}                          & \ldots \phantom{11} \\
G24A1 &   4$\times$15$^{15}$ & 1$\times$10$^{15}$  & 1$\times$10$^{16}$ &3$\times$10$^{15}$  \phantom{1}                            & 1$\times$10$^{16}$\phantom{11}  \\
G24A2 &  5$\times$10$^{15}$ & 2$\times$10$^{15}$ & 1$\times$10$^{16}$ &3$\times$10$^{15}$    \phantom{1}                            & 1$\times$10$^{16}$\phantom{11}   \\

\hline
\multicolumn{7}{|l|}{}. \\
\end{tabular}
\end{center}
$^{a}$ CH$_3$CN has been observed to be optically thick in all these objects \citep{beltran2005, beltran2011} so the column densities have been derived using observations of CH$_3$$^{13}$CN. \\
$^{\dagger\dagger}$ The column density of glycolaldehyde for G31 has been obtained using the rotation diagram method by \cite{beltran2009}\\
$^{\dagger\dagger\dagger}$ The column density for this object was determined using the rotation diagram method using a source size of 3\farcs5, see \ref{sec:mf}.
\end{table*}

Table \ref{tab:cd} shows that all our species peak in their density
in G31. However, it is surprising that there is relatively
little variation across the sample for HNCO and C$_2$H$_5$CN, both of which differ by a factor of 5 or
less from source to source. HNCO densities are remarkably consistent
throughout the sample. CH$_3$CN and CH$_2$(OH)CHO show more variation,
but if we exclude G31, then calculated column densities for the
remaining objects in the sample again agree very well, to within a
factor of $\sim$2.

\subsection{Spectral Modelling}
\label{sec:specmod}
\begin{figure*}
\begin{center}
\includegraphics[angle=-90,width=18cm, clip=true, trim=0cm 0cm 0cm 0cm]{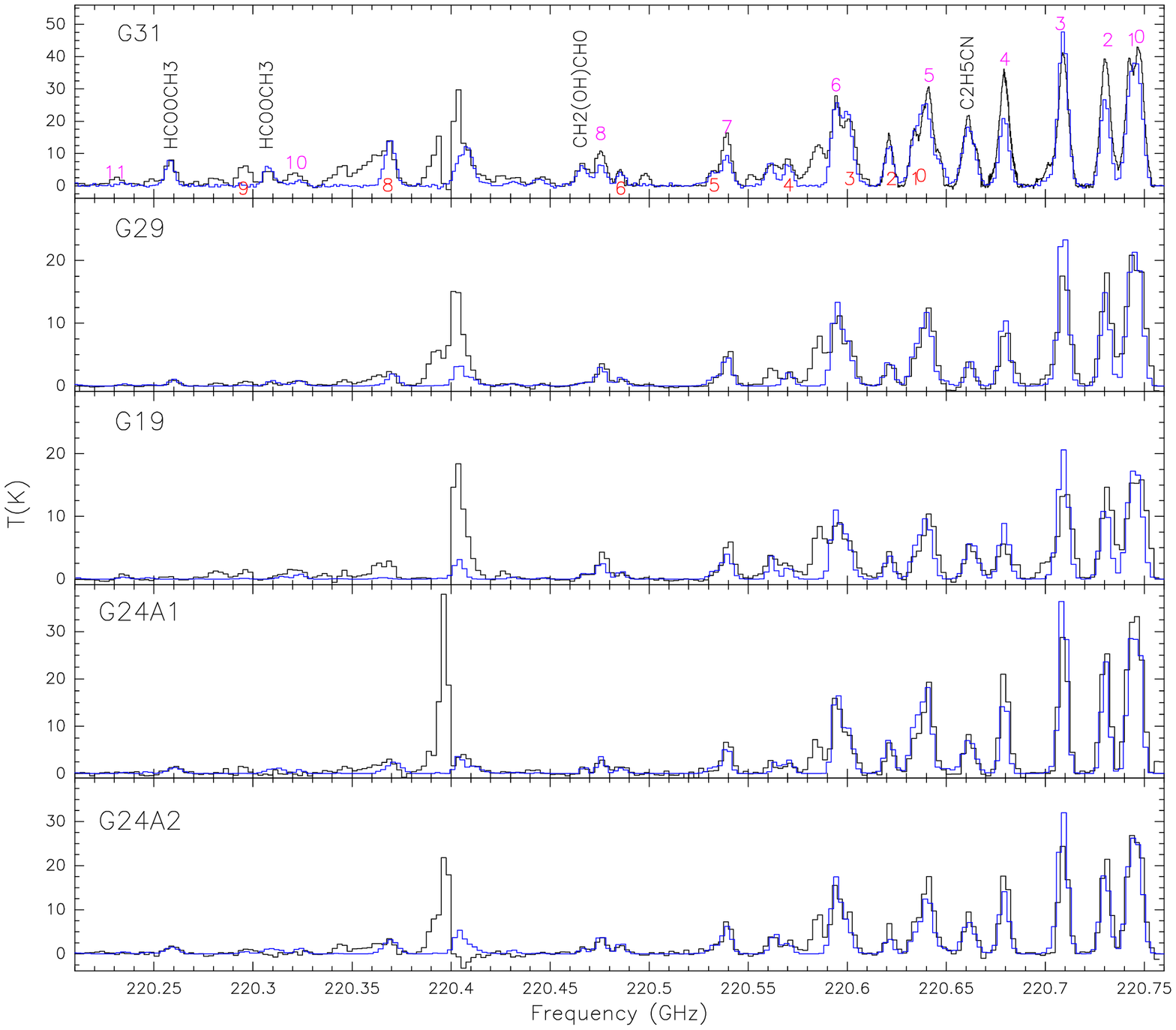}
\end{center}
\caption{CASSIS modelling (blue line) of methyl formate (HCOOCH$_{3}$), glycolaldehyde (CH$_2$(OH)CHO), methyl cyanide (CH$_3$CN), and ethyl cyanide (C$_2$H$_5$CN) overlaid on the PdBI observations (black line). We label the molecules detected in this work in the G31 panel. The numbers indicate the position of the CH$_3$CN~(12$_{\rm K}$--$11_{\rm K}$) K-components   in the upper part (in pink) of each spectra and of the  CH$_3$ $^{13}$CN~(12$_{\rm K}$--$11_{\rm K}$) K-components in the lower part (in red). We have excluded the $^{13}$CO line from these models since there is an issue of missing line flux in several of the hot core observations.  \label{fig:hcmod}}
\end{figure*}

We have also analysed the observations using the spectral modelling software CASSIS and using the JPL Catalog. CASSIS has been developed by IRAP-UPS/CNRS (http://cassis.irap.omp.eu). We use the LTE (local thermodynamic equilibrium) analysis tool to determine the column densities and excitation temperatures,  T$_{ex}$, required to reproduce the emission. The brightness temperature, T$_b$, of a given species is calculated by CASSIS according to:

\begin{equation}
T_{b}=T_{C}e^{-\tau} +(1-e^{-\tau})(J_{\nu}(T_{ex})- J_{\nu}(CMB))
\label{eq:tb}
\end{equation} 

where T$_{C}$ is the temperature of the continuum, $\tau$ is the opacity, CMB is the cosmic microwave background at 2.7 K, and  J$_{\nu}$(T) $ = h\nu/k)/(1-e^{h\nu/k T}-1)$ is the radiation temperature.

The input parameters for CASSIS are the column density, excitation temperature, source size, FWHM, and V$_{LSR}$ for each species we observe. Values for the source size, FWHM, and V$_{LSR}$ are taken from our observations. We vary the column densities and excitation temperatures for each species until a best fit is achieved. Further details of the CASSIS software and LTE analysis tool can found in the CASSIS documentation (http://cassis.irap.omp.eu/docs/RadiativeTransfer.pdf). \\

In particular, the spectral modelling focuses on the emission of CH$_3$CN, HCOOCH$_{3}$ and a possible contamination of the CH$_2$(OH)CHO emission with (CH$_3$)$_2$CO, to test the validity of our line assignments. In this model we assume the gas is in LTE conditions at a temperature T$_{ex}$. A comparison of the column densities we derive from observation and those we derive from spectral modelling can be found in Table~\ref{tab:specmod}.

Spectral models of CH$_3$CN emission, using only a single column density and excitation temperature, do not reproduce the observations accurately (Figure~\ref{fig:hcmod}). We hypothesise that this `poor' fit may be due to a combination of factors: (i) different transitions of CH$_3$CN may peak, or may be tracing, different temperatures and densities within our emission region, (ii) the emission we observe may not be in LTE, (iii) contributions due to blending, (iv) CH$_3$CN opacities are highly variable from the K=1 to the K=7 transitions, (v) the poor signal-noise ratio of the high energy K-transitions.  The best fit of all of the CH$_3$CN transitions is achieved at T$_{ex}$ ranging from 410 -- 450 K across our hot core sample. 
\begin{figure}
\begin{center}
\includegraphics[angle=-90,width=8cm, clip=true, trim=0cm 0cm 0cm 0cm]{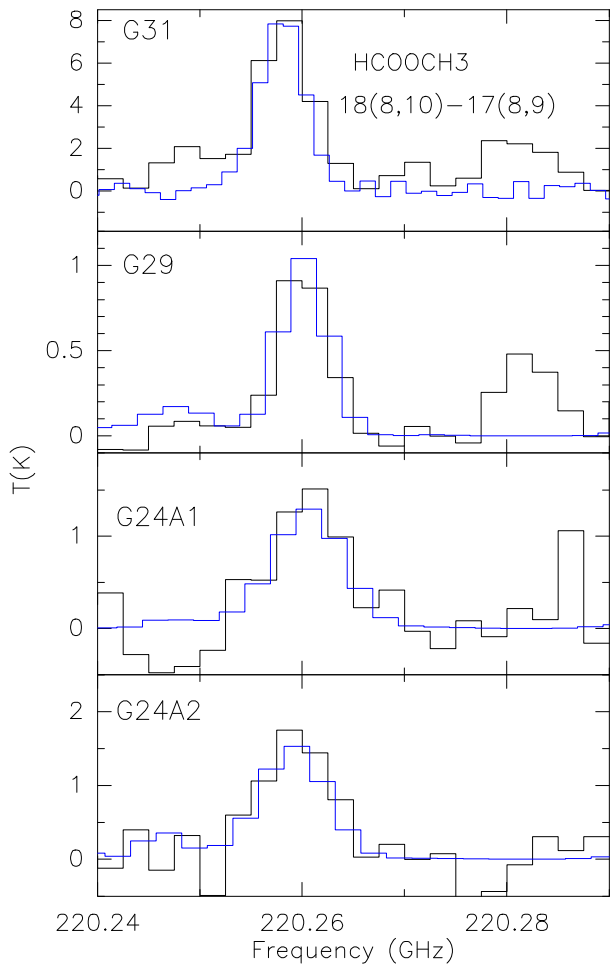}
\end{center}
\caption{Models of the $18_{8,10}$--$17_{8,9}$E transition of HCOOCH$_{3}$ using CASSIS (blue line) overlayed on the PdBI observations (black line).\label{fig:mfmod}}

\end{figure}
\begin{figure}
\begin{center}
\includegraphics[angle=-90,width=8cm, clip=true, trim=0cm 0cm 0cm 0cm]{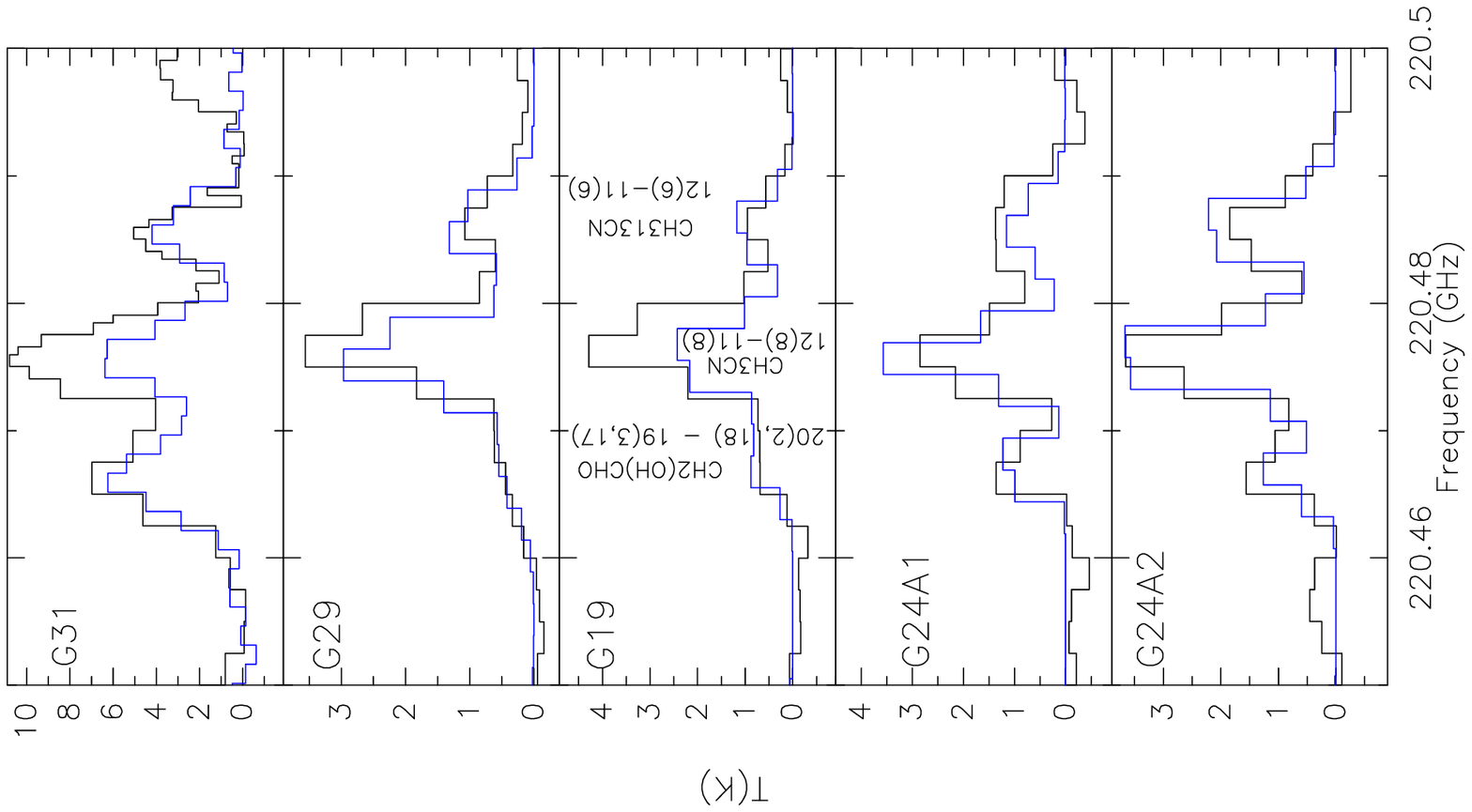}
\end{center}

\caption{CH$_2$(OH)CHO spectra with CASSIS models (blue line) overlaid on the PdBI observations (black line). From left to right: the 20$_{2,18}$-19$_{3,17}$ transition of glycolaldehyde, the 12(8)-11(8) transition of CH$_3$CN, and the 12(6)-11(6) transition of CH$_3$$^{13}$CN.\label{fig:glymod}}
\end{figure}
For HCOOCH$_{3}$ emission we have produced a spectral fit for the transition at 220.258 GHz (Figure~\ref{fig:mfmod}) and  the transition at 220.307 GHz. We are unable to model any isotopologues of methyl formate with CASSIS and have therefore omitted the transition at 220.553 GHz from our fit. In G31, where the observational column density was derived from multiple transitions of methyl formate, we find the required modelling column density to be the same as the observed value (4$\times$10$^{17}$ cm$^{-2}$). For the other hot cores we require a modelled column density a factor of 2 -- 3 larger than the observed value. This is not surprising as the observed column density was derived from only one transition in these cores and therefore represents a lower estimate of the methyl formate column density in these objects. As we only have one confirmed methyl formate transition in several of our hot cores, it is difficult to accurately measure the T$_{ex}$ required to reproduce these observations. 
We are confident from this modelling that we have correctly identified the $18_{8,10}$--$17_{8,9}$E  and  18$_{10,9}$--17$_{10,8}$E transitions of HCOOCH$_{3}$ in our hot core sample.\\

For CH$_2$(OH)CHO emission we have produced a spectral fit for the transition at 220.466 GHz (Figure~\ref{fig:glymod}). For G31 the required column density to reproduce the spectra is similar to the observed value. For G29, G19, G24A1, and G24A2 the required modelling column densities are a factor of 1.5 -- 3 larger than the observed values. This is likely due, again, to the observed column densities for these hot core being derived using only one CH$_2$(OH)CHO transition, and therefore represents a lower estimate of the column densities in these hot cores. \\

We have also explored the possibility that this transition of glycolaldehyde is blended with the 46$_{20,26}$--46$_{19,27}$ EE (E$_u$ = 816 K, S$\mu$$^2$ = 2843 D$^2$) and the 11$_{11, 1, 1}$ -- 10$_{10, 0, 1}$AE (E$_u$ = 63 K, S$\mu$$^2$ = 519 D$^2$)  transitions of acetone. We find for column densities and excitation temperatures of acetone which produce enough emission to explain the line at 220 466\,MHz, we overproduce emission at 220 368\,MHz. It is possible that both acetone and glycolaldehyde could be present in the hot cores in our sample, however, the contribution of acetone to the line seen at 220 466\,MHz is not significant ($<$7\%). More observations are needed to confirm the presence of both glycolaldehyde and acetone in these hot cores. \\

\begin{table*}
\caption{ Spectral modelling column densities and excitation temperatures of methyl cyanide, methyl formate and glycolaldehyde. \label{tab:specmod}}
\begin{center}
\footnotesize{
\begin{tabular}{cccc}
\hline
Hot Core&T$_\mathrm{ex}$ (K) & Modelled column density (cm$^{-2}$)& Observed at column density (cm$^{-2}$) \\
\hline
&&CH$_3$CN&\\
\hline
G31&450&9$\times$10$^{15}$&5$\times$10$^{16}$ (300 K)\\
G29&410&5$\times$10$^{15}$&2$\times$10$^{16}$ (300 K)\\
G19&450&4$\times$10$^{15}$&1$\times$10$^{16}$ (300 K)\\
G24a1&310&4$\times$10$^{15}$&4$\times$10$^{16}$ (300 K)\\
G24a2&450&6$\times$10$^{15}$&3$\times$10$^{16}$ (300 K)\\
\hline
          &       &  HCOOCH$_{3}$&\\
\hline
G31&130&4$\times$10$^{17}$&4$\times$10$^{17}$ (150 K)\\
G29&300&5$\times$10$^{16}$&2$\times$10$^{16}$ (300 K)\\
G24A1&300&9$\times$10$^{16}$&3$\times$10$^{16}$ (300 K)\\
G24A2&300&1$\times$10$^{17}$&4$\times$10$^{16}$ (300 K)\\
\hline
&&CH$_2$(OH)CHO &\\
\hline
G31&130&3$\times$10$^{16}$&2$\times$10$^{16}$ (150 K)\\
G29&300&8$\times$10$^{15}$&5$\times$10$^{15}$ (300 K)\\
G19&300&9$\times$10$^{15}$&3$\times$10$^{15}$(300 K)\\
G24A1&300&1$\times$10$^{16}$&5$\times$10$^{15}$ (300 K)\\
G24A2&300&1$\times$10$^{16}$&6$\times$10$^{15}$ (300 K)\\
\hline






\end{tabular}
}
\end{center}
\end{table*}

Chemical differentiation in hot cores has been observed before
\citep{mookerjea2007,garay1999} and attributed to either differences
in their host star mass and/or differences in ages. Results of
chemical models in the past have suggested that determining abundance ratios of typical hot core tracers (e.g. CH$_3$CN) as
well as sulphur-bearing species could serve as indicators of both mass
and age \citep{hatchell1998,viti2001,buckle2003,viti2004}. However much
care has to be taken due to the fact that most species do not
necessarily trace the very inner part of the core and their emission
region probably spans a range of densities and temperatures that make
the interpretation difficult \citep{wakelam2004}.  The complex species
we have been discussing in this paper, on the other hand, may provide
us with a better choice of age/mass discrimination as they
all seem  to trace a more compact region than the more
widely-observed species such as CH$_3$OH and CH$_3$CN. 
In the next section we make use of a chemical model, UCL\_CHEM
\citep{viti2004} to simulate the formation and evolution of hot cores
and see if the differences in ratios of the species listed in Table \ref{tab:cd}, between G31 and the rest of the cores, can shed some light on the masses and ages of our sources. 

\section{Chemical modelling}

UCL\_CHEM \citep{viti2004} is a two-phase time-dependent model which
follows the collapse of a prestellar core (Phase I), followed by the
subsequent warming and evaporation of grain mantles (Phase II). Phase
I starts from the number density of a diffuse cloud
(10$^2$\,cm$^{-3}$) and allows a free-fall collapse to take place
until a final density (which varies between 10$^{6-8}$\,cm$^{-3}$) is
reached.  This occurs isothermally at a temperature of 10\,K. During
the collapse, atoms and molecules collide with, and freeze on to,
grain surfaces.  The depletion efficiency is determined by the
fraction of the gas-phase material that is frozen on to the grains
\citep{rawlings1992}. This fraction is arrived at by adjusting the grain
surface area per unit volume, and assuming a sticking probability of
unity for all species. The fraction of material on grains is then
dependent on the product of the sticking probability and the amount of
cross-section provided per unit volume by the adopted grain size
distribution. Grains are considered to be spheres.  We assume that
hydrogenation occurs rapidly on the grain surfaces, so that, for
example, some percentage of carbon atoms accreting will rapidly become
frozen out methane (CH$_4$) etc. In Phase II we increase the dust and gas temperature up to 300 K, to simulate the presence of a nearby infrared source in the core. As the temperature increases, mantle species desorb in various temperature bands (see \citet{collings2004}). UCL\_CHEM treatment of the temperature and of the ice sublimation is as in \citep{viti2004}, where details of how the temperature increases with time leading to the subsequent time dependent sublimation of the ice mantles can be found. 

Initial atomic abundances are taken from \citet{sofia2001}, as in
\citet{viti2004}. For the gas-phase chemistry, reaction rate
coefficients are taken from the UMIST database \citep{woodall2007}.
Some coefficients have been updated with those from the KIDA database
\citep{wakelam2009}. Our database also includes some simple
grain-surface reactions (mainly hydrogenation) as in \citet{viti2004}
as well as selected routes for grain-surface formation of
glycolaldehyde as in \citet{woods2012} and methyl formate as in
\citet{occhiogrosso2011}. In Phase I non-thermal desorption is
considered as in \citet{roberts2007}.

%
%
%
%

We investigate how varying key model parameters, such as the final
collapse density, lifetime of the cold phase, the type of evaporation,
and the efficiency of the freeze-out of species (measured as a
percentage of the total CO in the solid phase) affect
the abundances of methyl cyanide, methyl formate, glycolaldehyde,
ethyl cyanide and isocyanic acid during the hot core evolution.  We
have modelled hot cores with final densities of $10^6$\,cm$^{-3}$,
$10^7$\,cm$^{-3}$ and $10^8$\,cm$^{-3}$. We also vary the efficiency
of the freeze-out of species from 14\%--100\%.  Since the period in
which the grains are warmed from very low temperatures to the
temperatures observed in typical hot cores is determined by the time
taken for a pre-stellar core to evolve towards the Main Sequence, and
hence by its mass \citep{viti1999} we have explored the effect of
new-born stars with different masses (15\,M$_\odot$ and
25\,M$_\odot$), corresponding to contraction times of 118\,000\,yr and
70\,000\,yr respectively -- see \citet{bernasconi1996}.

In the following section we explore the sensitivity of the
aforementioned species to changes in the physical and chemical
parameters described above, before comparing our theoretical models
with observations. Plots \ref{fig:mass_sens}--\ref{fig:abundvstime_fr0,18_glycole-17_mass15vap1} show
gas-phase abundances in Phase II of the model, for ease of comparison
to observationally-derived results.

\subsection{Sensitivity to stellar masses}

\begin{figure}
\begin{center}
\includegraphics[angle=-0,width=9cm, clip=true, trim=0.2cm 0cm 0cm 0cm]{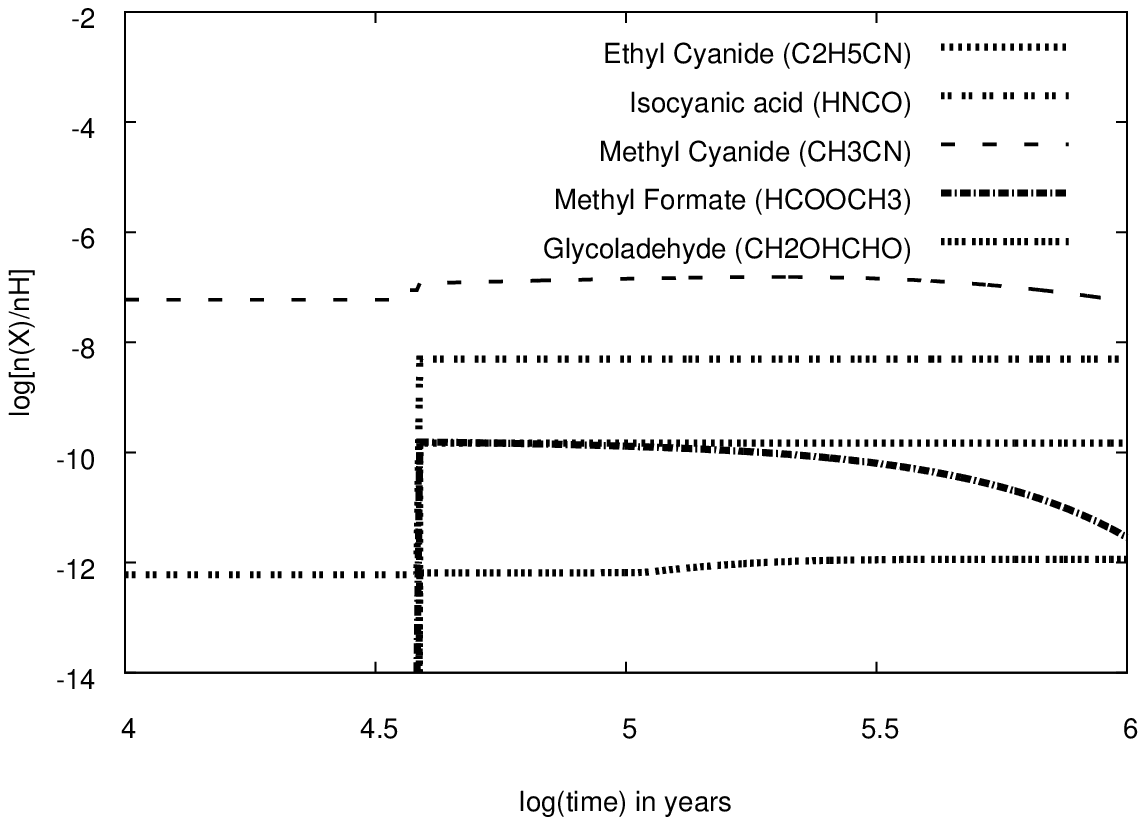}
\includegraphics[angle=-0,width=9cm, clip=true, trim=0.2cm 0cm 0cm 0cm]{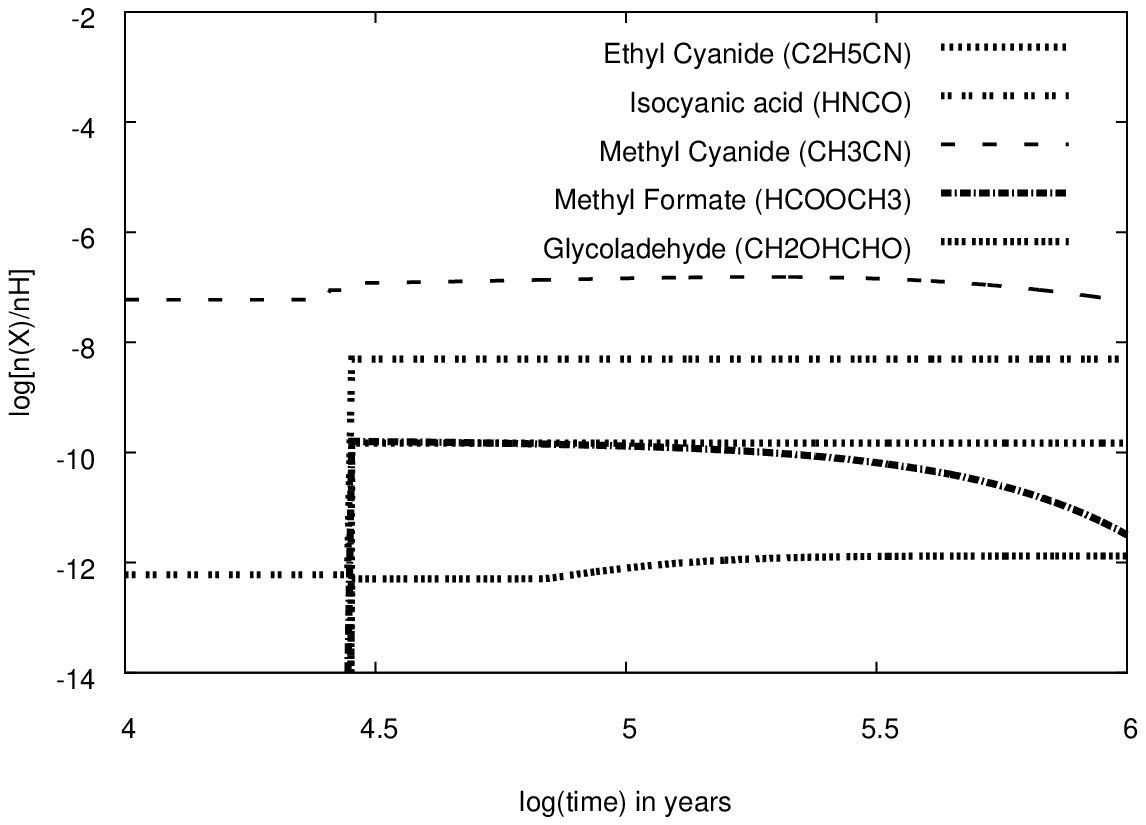}

\caption{Gas phase fractional abundances of selected species as a function of
  time for a 15\,M$_\odot$ (top) and a 25\,M$_\odot$ star (bottom),
  both with a final gas density after Phase II of
  10$^7$\,cm$^{-3}$.   At $\approx$ 10$^{4.5}$ years the sudden jump in abundance of methyl formate and ethyl cyanide is due to the evaporation of these species from the grains mantle surface.} \label{fig:mass_sens}
\end{center}
\end{figure}

Figure~\ref{fig:mass_sens} shows an example of two models, each with a
final density at the end of Phase I of 10$^7$\,cm$^{-2}$ and a
freeze-out percentage of $\approx$60\% in the solid phase, differing in the final mass of the star
in Phase II. At $\approx$ 10$^{4.5}$ years the sudden jump in abundance of methyl formate and ethyl cyanide is due to the evaporation of these species from the grain mantle surface. At late times varying the final mass of the star does not
seem to affect the fractional abundances of the complex species we
consider; however, as expected \citep[see][]{viti2004}, a higher mass
implies an earlier icy mantle evaporation. In general a very low
abundance of COMs implies a young age with the exception of methyl
cyanide (but see Sect.~\ref{sec:sensgasdens}) whose abundance is high
regardless of the age of the core; hence, a core where this species is
the {\it only} abundant complex molecule may be a very young core. The
theoretical low fractional abundances before evaporation
translate into column densities of the order of
$<$10$^{10}$\,cm$^{-2}$, much lower than the lowest observed values
for our sample.

\subsection{Sensitivity to the icy mantle composition}

\begin{figure}
\begin{center}
\includegraphics[angle=-0,width=9cm, clip=true, trim=0.2cm 0cm 0cm 0cm]{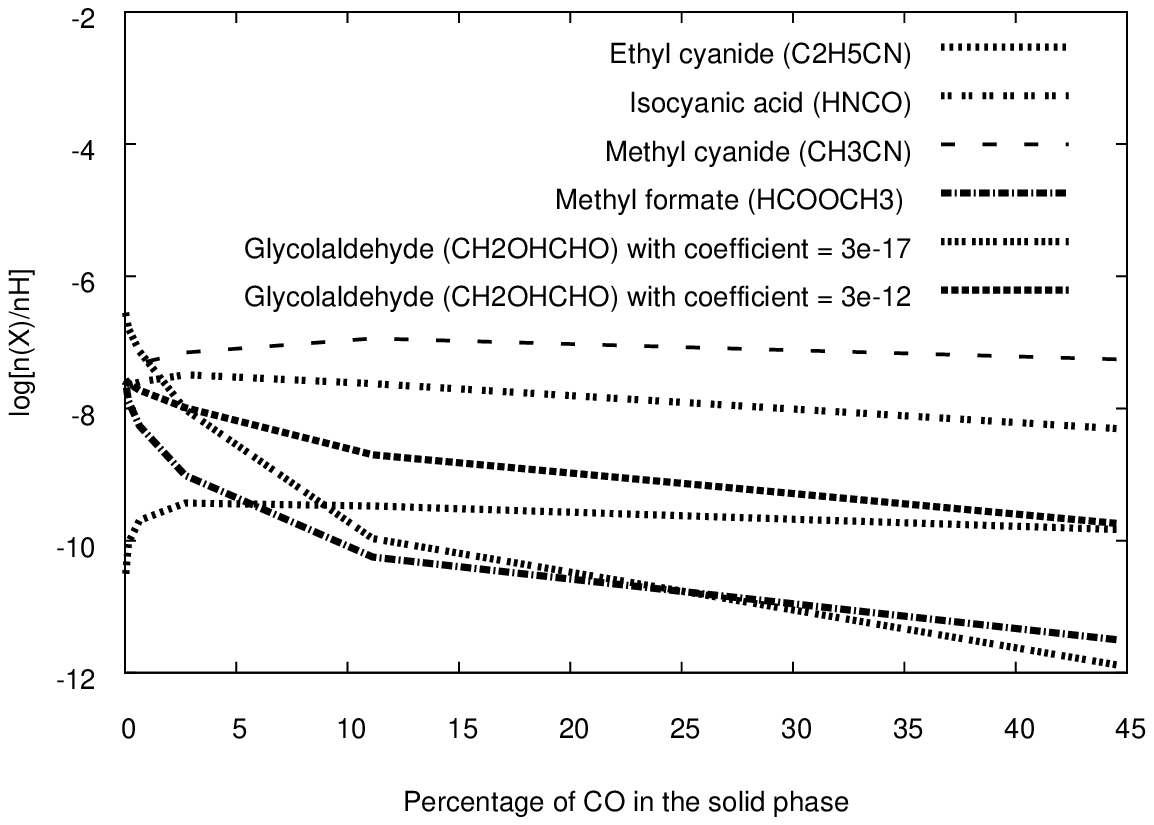}
\caption{Fractional abundances of selected species at 10$^6$\,yr after
  the `switch on' of the star as a function of the percentage of CO left in
  the solid phase at the end of Phase I, for a 25\,M$_\odot$ star with a
  final gas density after phase I of
  10$^7$\,cm$^{-3}$. \label{fig:mass_fr}}
\end{center}
\end{figure}

Figure~\ref{fig:mass_fr} is a plot of the fractional abundances of
selected species at 10$^5$\,yr after Phase II starts, as a function of
the percentage of CO left in the solid phase at the end of Phase I. Note that
after few thousand years from the beginning of Phase II CO will have
completely evaporated back into the gas phase so the percentage on the
$x$-axis is {\it not} an indication of the final gas CO abundance.  In the figure, we present models which use two different values of the reaction rate coefficient for the formation of glycolaldehyde. This is due to the uncertainties in the value of this quantity (see \citet{woods2012}). Based on the sensitivity analysis of \citep{woods2012}, we only consider one formation route for glycolaldehyde for simplicity: g-CH$_3$OH + g-HCO $\rightarrow$  g-CH$_2$(OH)CHO, where g- refers to a species which is frozen onto the grain surface.      
As expected, the abundance of most complex species increases with an
increase in freeze-out efficiency (corresponding to a decrease in
percentage of CO in the gas phase). However it is interesting to note
that ethyl cyanide seems to decrease at a very high freeze out
efficiency while the abundance of methyl cyanide is more or less
constant. The main route of formation for ethyl cyanide is via
hydrogenation of HC$_3$N on the
grains so in principle its abundance should increase with the
efficiency of freeze-out. However, as HC$_3$N  is efficiently formed in the gas phase via reactions
involving H$_2$O, HCN and other hydrocarbons, when
freeze-out is very efficient the reactants are depleted from gas deposited on the grains
at a higher efficiency than that of the reactions forming HC$_3$N.

\subsection{Sensitivity to the gas density}
\label{sec:sensgasdens}

\begin{figure}
\begin{center}
\includegraphics[angle=0,width=9cm, clip=true, trim=0.2cm 0cm 0cm 0cm]{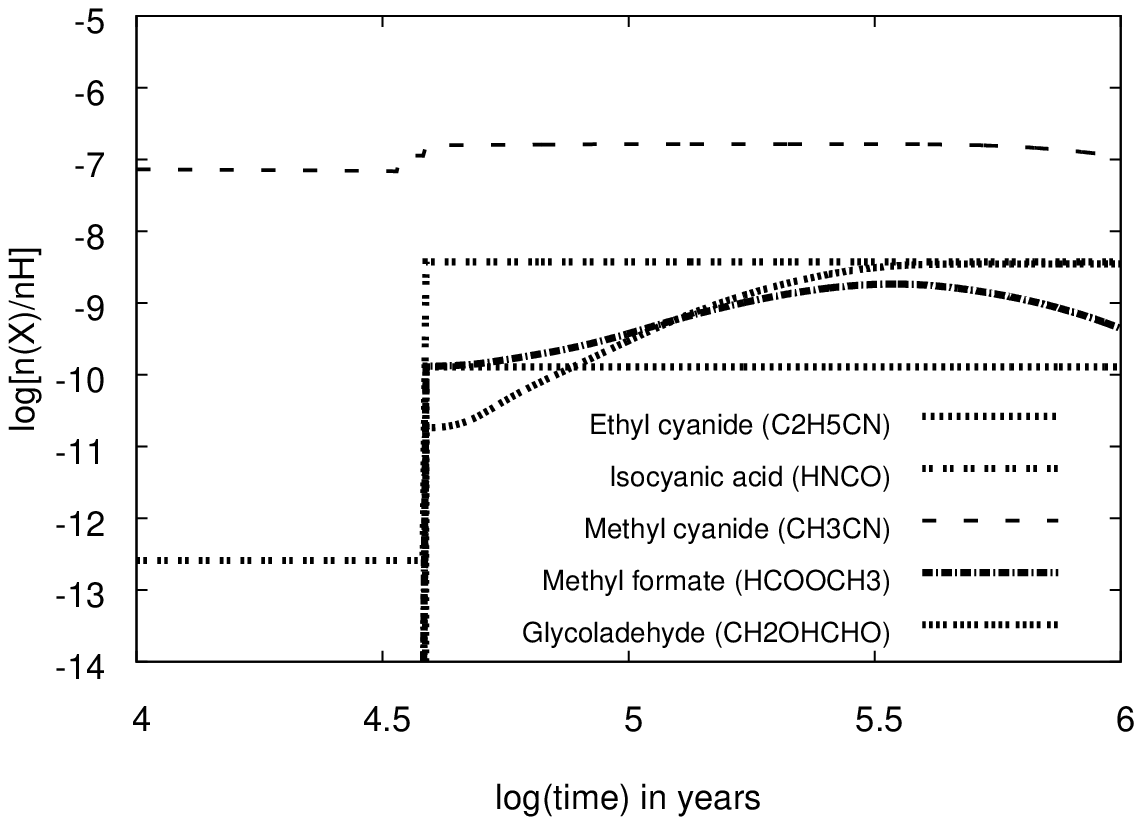}
\includegraphics[angle=0,width=9cm, clip=true, trim=0cm 0cm 0cm 0cm]{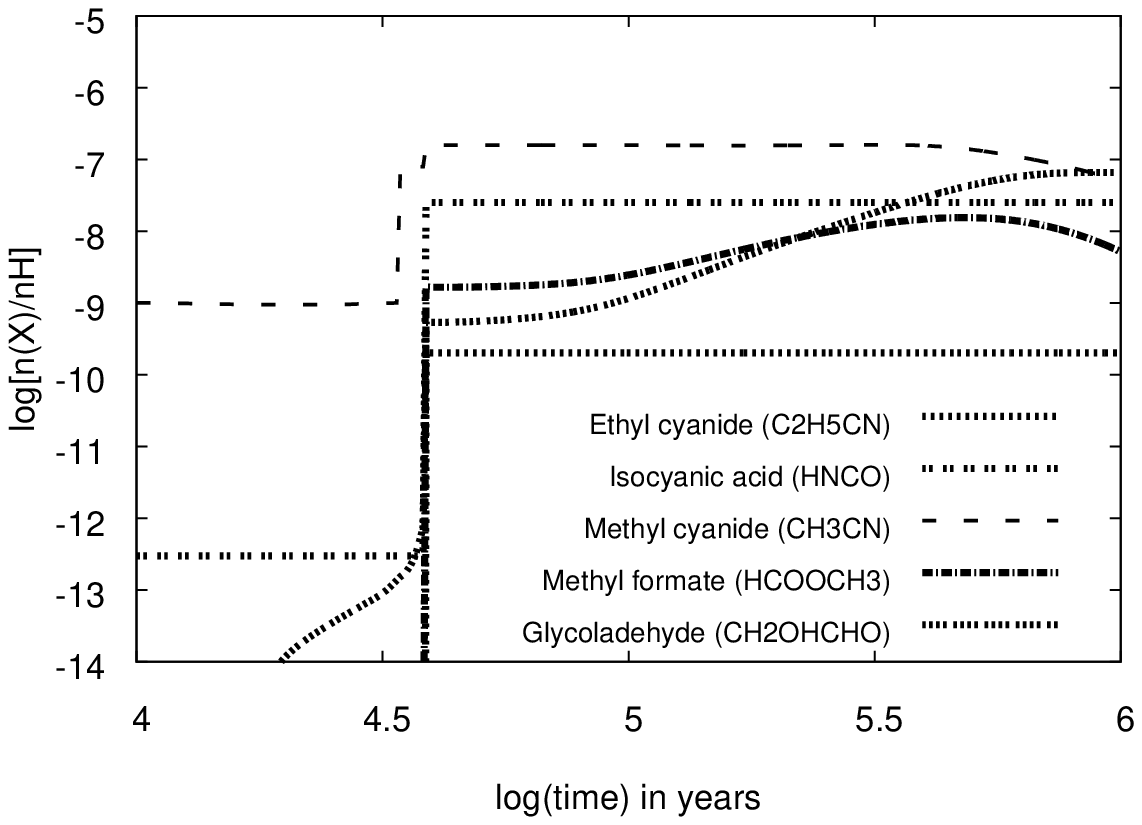}
\includegraphics[angle=-0,width=9cm, clip=true, trim=0cm 0cm 0cm 0cm]{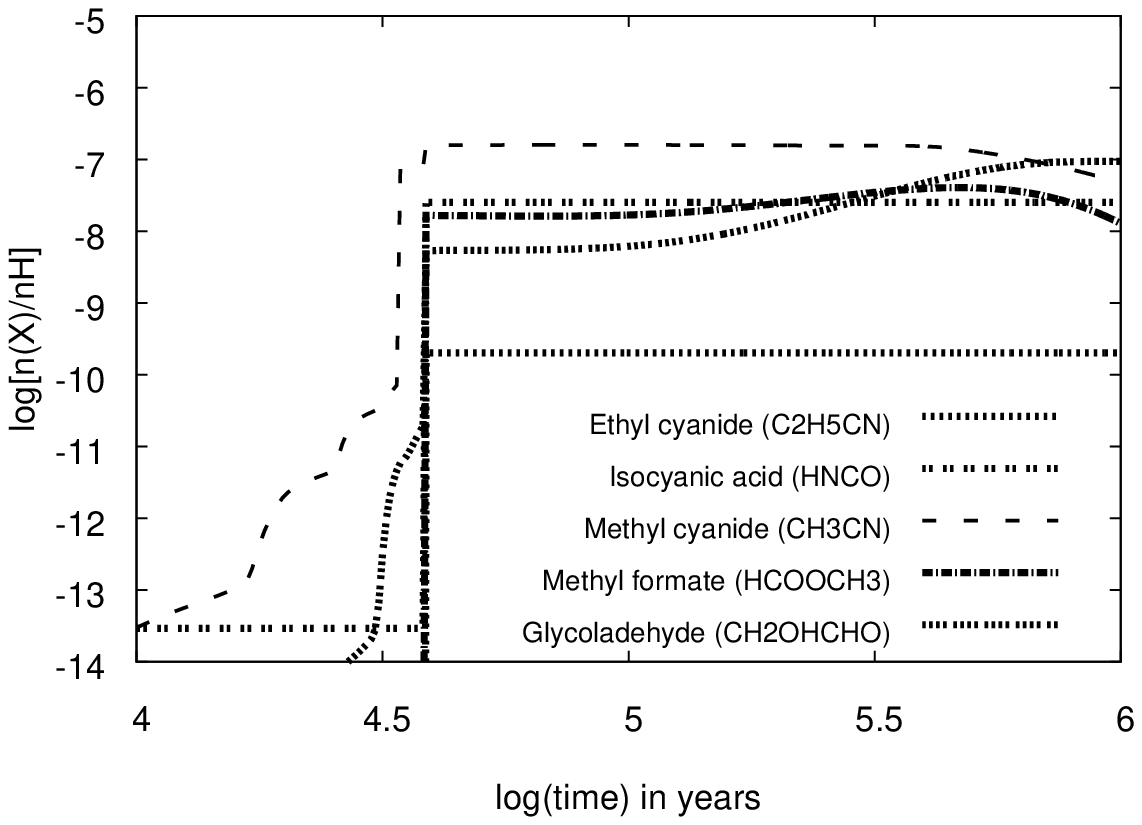}

\caption{The gas phase abundance of isocyanic acid, ethyl cyanide, methyl
  cyanide, methyl formate, and glycolaldehyde as a function of time
  for the percentage of CO left in the solid phase of 100\% , a mass of 15\,M$_\odot$, equivalent to a
  contraction time of 115\,000\,yr, with step evaporation of the icy
  mantles.  The top panel corresponds to a final density after phase I
  of 10$^6$\,cm$^{-3}$. The middle panel corresponds to a final
  density after phase I of 10$^7$\,cm$^{-3}$. The bottom panel
  corresponds to a final density after phase I of
  10$^8$\,cm$^{-3}$. \label{fig:abundvstime_fr0,18_glycole-17_mass15vap1}}
\end{center}
\end{figure}

Figure~\ref{fig:abundvstime_fr0,18_glycole-17_mass15vap1} shows the
abundance of our selected molecules as a function of time for
three different final densities. The fractional abundance of most species
increases with density when going from 10$^6$ to 10$^7$\,cm$^{-3}$
although the increase is less pronounced at high densities. An
interesting effect of increasing densities is observed with methyl
cyanide whose abundance before thermal evaporation is still high at
the lowest density while it drops considerably at high densities: we
interpret this as a direct effect of freeze-out, which is directly
proportional to collisional frequency of the parent species forming
CH$_3$CN during phase I. Finally we note that, at a late stage of the
evolution of the core, the ratio of the selected COMs varies as a
function of density: for example, at lower densities CH$_3$CN is
always higher than the other organic molecules while already at
10$^7$\,cm$^{-3}$ CH$_3$CN/C$_2$H$_5$CN is $\sim$1 and at higher
densities, $<$1. Similar considerations can be made about other
species, making the ratios of our selected COMs ideal tracers of
densities for evolved hot cores.

\subsection{Comparison with observations}

Table \ref{tab:cd} shows that the differences in column densities
among cores, for most species, are seldom larger than one order of
magnitude; considering the uncertainties in model parameters as well as formation and destruction
rates for COMs, it is therefore not possible, at present, to use chemical models
to infer the age or the mass of each individual core. Nevertheless,
we do attempt a qualitative comparison with
observations by estimating
theoretical column densities from our models. We derive the theoretical column
density $N$ by using the approximate formula below:
\begin{equation}
N = X \times A_\mathrm{V} \times \mathrm{N}_{H2},
\end{equation}
where $X$ is the fractional abundance from our models and N$_{H2}$ is the column density of H$_2$ that provides 1 mag of extinction \citep{bohlin1978}, which is 1.6$\times$10$^{21}$ cm$^{-2}$. For
$A_\mathrm{V}$ we adopt a typical hot core visual extinction of
600\,mags  (noting that the column density simply
scales linearly with extinction and that the fractional abundances
calculated from the models are insensitive to visual extinctions above
a critical value of $\sim$10\,mags, when photons do not penetrate any
further).

The first general conclusion we can draw by comparing Table \ref{tab:cd} with our models (see Figs. 13, 15 and 16) is that all cores are evolved enough that all the mantles have evaporated. From our models this means that they are at least 20,000 yr old cores (this estimate increases with decreasing mass).

We find that most models match the theoretical abundance of HNCO quite well: the theoretical
column densities vary between 8$\times$10$^{15}$ and 3$\times$10$^{16}$ cm$^{-2}$ where the upper values coincide with models with a high density and freeze out (as expected since HNCO forms mainly on the grains). HNCO seems to be fairly constant among cores and its observed column densities are closer to the upper theoretical values;  hence the only tentative conclusion we can derive here is that the density of the observed cores is $>$10$^6$ cm$^{-3}$, possibly as high as 10$^{8}$ cm$^{-3}$; this is consistent with the emission of COMs being so compact that they trace the gas closer to the protostar.

C$_2$H$_5$CN is, on the other hand, always under abundant in our models although, again, it is highest in models where the density and freeze out are high. We note that, among the cores, the highest value for this species is found in G31, implying a higher density for this core.

CH$_3$CN shows a larger variation in observed column densities, from 10$^{16}$ cm$^{-2}$ in G19 to 5$\times$10$^{16}$ in G31. A similar range is found in our models as a function, again, of density, although interestingly not of freeze out (see Fig. 14). Again we conclude that G31 is the densest core, with $n_{H}$ close to 10$^8$ cm$^{-3}$.

The  observed CH$_2$(OH)CHO column density is similar among all cores but G31, where it is two orders of magnitude higher than in the other objects.  It is interesting to note that we can match its column density with models of gas densities of the order of 10$^8$ cm$^{-3}$ for G31 and 10$^6$--10$^7$ cm$^{-3}$ for the rest of the cores, again supporting the conclusion that G31 has a higher gas densities than the rest of the objects in our sample.   We note that only models where we use a high rate coefficient for the formation of glycolaldehyde are able to reproduce the observations, in agreement with the findings of \citep{woods2012}.

Finally, HCOOCH$_3$ shows the same behaviour as glycolaldehyde in that it is higher by at least one order of magnitude in G31 compared to the rest of the sample where it is more or less constant at 10$^{16}$ cm$^{-2}$. In our models, we reach a  column density of 10$^{16}$ cm$^{-2}$ at gas densities of 10$^7$ cm$^{-3}$; interestingly an increase to 10$^8$ cm$^{-3}$ only yields an increase in methyl formate by a factor of few and only for a relatively short period of time, since this species seems to decline in abundance after 5$\times$10$^5$ yrs. However we point out that the column density for G31 was derived using a temperature derived from a rotational diagram, while for all the other cores a simple LTE calculation at 300K was performed.
In fact, in general, LTE calculations at lower temperature would yield a lower column densities.

\section{Conclusions}
We have analysed IRAM PdBI data, in the frequency range  220\,209.95\,MHz to 220\,759.69\,MHz, towards six hot cores: G31.41+0.31, G29.96-0.02, G19.61-0.23, G10.62-0.38, G24.78+0.08A1 and G24.78+0.08A2. The aim was to identify seven lines that were unidentified by \citet{beltran2005} in G31 and look for their presence in the other five hot cores, as well as identify other complex molecules that were identified by \citet{beltran2005} in G31 and G24 but not in G29, G19 and G10.\\

We have identified three new transitions of methyl formate (HCOOCH3) in G31, two of which are vibrationally excited lines. A rotation diagram analysis of these lines combined with those in \citep{fontani2007} yields a column density for methyl formate of 4$\times$10$^{17}$ cm$^{-2}$. This is at least two orders of magnitude larger than the column densities in the other hot cores in our sample. We have also found a single temperature component of $\sim$170 K (using a source size of 3\farcs5) that fits these transitions, although the vibrationally-excited transitions we have found and our very large column densities would suggest that methyl formate may trace multiple temperature components of G31. This may also be true of the hot cores in G29 and G24. The spatial distribution of methyl formate does indicate that it traces
the dense and compact parts of hot cores. Comparatively methyl formate traces a region slightly more compact than that of methyl cyanide but glycolaldehyde emission still remains the most compact to date. At this stage we can not conclude whether the fact that methyl formate is more extended than glycolaldehyde  is a question of excitation or chemistry. We do find, however, that in G29 methyl formate and glycolaldehyde are tracing a an emission region of 0.05 pc which is comparable to the compact emission of glycolaldehyde in G31 (0.08pc). In G24 methyl formate and glycolaldehyde both trace a region comparable in size to the region traced by methyl formate in G31. In our models we do not find a density where there is more methyl formate than glycolaldehyde which would suggest that glycolaldehyde forms in a smaller denser region. We note, however our glycolaldehyde network of reactions is far from complete and it is therefore likely that we form too much glycolaldehyde in our model.\\

In this work we postulate that we have found the 20$_{2,18}$-19$_{3,17}$ transition of glycolaldehyde in two more hot cores bringing the total number of detections in high mass star forming regions, outside the Galactic Centre, to five hot cores.  This emission whilst being far more compact in G31, is of comparable compactness to methyl formate in G24A1, G24A2 and G29. We have spectrally modelled our emission to explore the possible contamination of this transition with the 46$_{20,26}$--46$_{19,27}$ EE and 11$_{11, 1, 1}$ -- 10$_{10, 0, 1}$AE transitions of acetone. We find that any contribution of acetone emission to the lie seen at 220 466\,MHz is not significant. More observations are needed to confirm the presence of glycolaldehyde in these hot cores.  Our other complex molecule detections in our sample highlight chemical homogeneity among G29, G19, G24A1 and G24A2, not only in terms of presence or absence of certain transitions but also when comparing column densities. G31, however, is the most chemically rich object and the significantly  different column densities we find in this core and the variety of transitions seen may suggest that it represents a different evolutionary stage to the other hot cores in our sample, or it may surround a star with a higher mass. \\

We have also undertaken a comparison between observations and a chemical model, UCL CHEM, to interpret the molecular inventory of the six cores and qualitatively characterise each core and its evolutionary stage. We note that of the species we are modelling, only methyl formate and methyl cyanide have been extensively studied in the laboratory \citep{modica2010, bennett2007, khlifi1996, defrees1985, huntress1979}. The other complex molecules are little-known and it is very likely that our models are missing routes of formation and destruction for these species. Moreover, we note that the LTE calculations shown in Table 6 were for a temperature of 300K; a lower temperature would in general yield a lower column density. The uncertainties related to the size of the emission region and temperatures, together with the incompleteness of the chemical networks for COMs makes a more quantitative comparison with chemical modelling inappropriate. \\

In conclusion, a qualitative comparison between our modelling and observations seem to consistently yield a higher density for G31 than the other objects in our sample, a result consistent with the fact that most lines are indeed the brightest in G31. We can also safely conclude that our sample only contains evolved hot cores, with an age of at least 20,000 years. We are unable to constrain the mass of each core; this information would have led to a better constraint for the age of each core.

\section*{Acknowledgments}
This research is supported by an STFC PhD studentship and the LASSIE Initial Training Network under the European Community’s Seventh Framework Programme FP7/2007-2013 under grant agreement No. 238258. Funding for this work was also provided by the Leverhulme Trust.

Data was reduced using the Gildas software package \footnote{http://www.iram.fr/IRAMFR/GILDAS}.
Spectral line data were taken from the Spectral Line Atlas of
Interstellar Molecules \citep[SLAIM; Available at
  \url{http://www.splatalogue.net}. F. J. Lovas, private
  communication,][]{remijan2007}, the JPL Spectral Line Catalog
\citep{pickett1998}, the Cologne Database for Molecular Spectroscopy
\citep{muller2005} and the Lovas/NIST database \citep{lovas2004}.

\bibliography{linespaperVref3}
\bibliographystyle{mn2e}

\label{lastpage}

\end{document}